\documentclass[11pt, a4paper, twoside]{article}
 \usepackage[font=small,format=plain,labelfont=bf,up,textfont=normal,up,justification=justified,singlelinecheck=false]{caption}

\usepackage[a4paper, left=3cm,right=2cm]{geometry}
\usepackage{hyperref}
\usepackage{amsfonts}
\usepackage{amsmath}
\usepackage{setspace}
\usepackage{color}
\usepackage{multirow}
\usepackage[]{youngtab}
\usepackage[utf8,applemac]{inputenc}
\usepackage{tensor}
\usepackage{cite}
\usepackage{tikz}
\usepackage{graphicx}
\graphicspath{{figure/}}
\usepackage{slashed}

\usepackage{dcolumn}
\usepackage{bm}
\usepackage[justification=centering]{caption}

\newcommand{\bea}{\begin{eqnarray}}
\newcommand{\eea}{\end{eqnarray}}
\newcommand{\be}{\begin{equation}}
\newcommand{\ee}{\end{equation}}
\def\nn{\nonumber}

\def\p{\partial}

\newcommand{\cR}{\mathcal{R}}
\newcommand{\cC}{\mathcal{C}}
\newcommand{\cD}{\mathcal{D}}
\newcommand{\cO}{\mathcal{O}}
\newcommand{\cT}{\mathcal{T}}

\newcommand{\Tr}{\textrm{Tr}}
\newcommand{\tr}{\textrm{tr}}

  \newcommand{\beqs}{\begin{eqnarray}}
\newcommand{\eeqs}{\end{eqnarray}}

\def \D {\Delta}

\def \m {\mu}

\def \r {\rho}

\def \th {\theta}

\def \p {\partial}
\def \f {\frac}

\def \nn {\nonumber}
\def \hs {\hspace}
\def \la {\langle}
\def \ra {\rangle}

\begin{document}
\title{On the Mutual Information in Conformal Field Theory}
\author{
Bin Chen$^{1,2,3}$,\footnote{bchen01@pku.edu.cn}\,
Lin Chen$^{1}$,\footnote{1401110063@pku.edu.cn}\,
Peng-xiang Hao$^{1}$\footnote{pxhao@pku.edu.cn}\,
and
Jiang Long$^{4}$\footnote{Jiang.Long@ulb.ac.be}
}
\date{}

\maketitle

\begin{center}
{\it
$^{1}$Department of Physics and State Key Laboratory of Nuclear Physics and Technology,\\Peking University, 5 Yiheyuan Rd, Beijing 100871, P.R.\,China\\
\vspace{2mm}
$^{2}$Collaborative Innovation Center of Quantum Matter, 5 Yiheyuan Rd, \\Beijing 100871, P.R.\,China\\ \vspace{1mm}
$^{3}$Center for High Energy Physics, Peking University, 5 Yiheyuan Rd, \\Beijing 100871, P.R.\,China\\ \vspace{1mm}
$^{4}$Universit\'{e} Libre de Bruxelles and International Solvay Institutes,\\
CP 231, B-1050 Brussels, Belgium}
\vspace{10mm}
\end{center}

\begin{abstract}

In this work, we study the universal behaviors in the mutual information of two disjoint spheres in a conformal field theory(CFT). By using the operator product expansion of the spherical twist operator in terms of the conformal family, we show that the large distance expansion of the mutual information can be cast in terms of the conformal blocks. We develop the $1/n$ prescription to compute the coefficients before the conformal blocks. For a single conformal family, the leading nonvanishing contribution to the mutual information comes from the bilinear operators. We show that the coefficients of these operators take universal forms and  such universal behavior persists in the bilinear operators with derivatives as well. Consequently the first few leading order contributions to the mutual information in CFT take universal forms. To illustrate our framework, we discuss the free scalars and free fermions in various dimensions. For the free scalars, we compute the mutual information to the next-to-leading order and find good agreement with the improved numerical lattice result. For the free fermion, we compute the leading order result, which is of universal form, and find the good match with the numerical study. Our formalism could be applied to any CFT potentially.

\end{abstract}
\baselineskip 18pt
\thispagestyle{empty}

\newpage

\section{Introduction}

Entanglement entropy encodes the hidden information of a region with its environment. It is defined to be the von Neumann entropy of the reduced density matrix\cite{nielsen2010quantum,petz2008quantum}
\be
S_A=-\Tr_A \r_A\log \r_A, \hs{5ex}\r_A=\tr_{\bar A} \r,
\ee
where $\r$ is the density matrix with respect to the wavefunction of the whole system. The entanglement entropy has been an important quantity in many body quantum systems, for example as the order parameter in quantum phase transition. Very recently it was found to be at the crossing point of quantum gravity, quantum field theory and quantum information. Especially following the proposal of the holographic entanglement entropy\cite{Ryu:2006bv,Ryu:2006ef}, it has opened a new window to study the AdS/CFT correspondence and the holographic principle\cite{Maldacena:1997re}.

The entanglement entropy in a quantum field theory is hard to compute due to the infinite degrees of freedom in a field theory. In the seminal paper \cite{Srednicki:1993im}, it was found that the entanglement entropy actually obeys an area law, being proportional to the area of the boundary of the entangle region\cite{Eisert:2008ur}. The usual way to compute the entanglement entropy in a quantum field theory is to apply the replica trick to define the R\'enyi entanglement entropy\cite{Holzhey:1994we}
\be
S_A^{(n)}=-\f{1}{n-1} \log \Tr_A \r_A^n.
\ee
It is easy to see that the entanglement entropy can be obtained by taking the $n\to 1$ limit of the R\'enyi entropy
\be
S_A=\lim_{n \to 1} S_A^{(n)}, \label{limit}
\ee
provided that such a limit is well-defined. However, the replica trick requires computing  the partition function of a multi-sheeted space or introducing the twist fields to
 impose nontrivial monodromy condition on the fields\cite{Calabrese:2004eu,Calabrese:2005zw}. Another way to compute the entanglement entropy in the free field theory is to apply an algebraic way\cite{Casini:2009sr}. Most of the study of the entanglement entropy in quantum field theory has been focused on the simplest entangling region, the half space or a single sphere, in the free theory and  conformal field theory(CFT).

The study of the entanglement entropy in CFTs has drown much interest in the past decade due to its relation with holography and quantum gravity\cite{Ryu:2006bv,Ryu:2006ef}.
The conformal symmetry in a CFT often allows us to get more exact results. For example,
 the single-interval entanglement and R\'enyi entropies in a two-dimensional conformal field theory (CFT) display a universal behavior, being proportional to the central charge of the CFT\cite{Holzhey:1994we,Calabrese:2004eu}. However such universal behavior is rare in higher dimensional CFTs. In this work we would like to report the universal behaviors in the mutual information in any CFTs.

The mutual information is defined with respect to two subsystems
 $A$ and $B$ as
\be
I_{A,B}=S_{A}+S_{B}-S_{A\cup B}.
\ee
The mutual information of two disjoint regions $A$ and $B$ is interesting.  First of all, it measures the quantum correlation between two regions. It is  positive, finite, free of ultra-violet divergence. In particular the mutual information satisfies the  inequality\cite{Wolf}
\be
I_{A,B}\geq \frac{\cC(M_A,M_B)^2}{2\parallel M_A\parallel^2 \parallel M_B\parallel^2},
\ee
where $M_A$ and $M_B$ are the observables in the regions $A$ and $B$ respectively, and $\cC(M_A,M_B):=\la M_A\otimes M_B \ra-\la M_A \ra \la M_B\ra $ is the connected correlation function of $M_A$ and $M_B$. Holographically the classical mutual information of two well separated regions is vanishing, but the above inequality suggests that the quantum correction should give nonvanishing contribution\cite{Faulkner:2013ana}. In AdS$_3$/CFT$_2$, if one considers two intervals far apart  in the vacuum state of the large $c$ CFT, the leading contribution in the mutual information is vanishing but the subleading term independent of $c$ is not vanishing, which is in perfect match with 1-loop partition function of the graviton\cite{Barrella:2013wja, Chen:2013kpa}.

It is a formidable problem to compute  the mutual information of two disjoint regions directly via the  replica trick. The replicated geometry is not only of singularity but also of nontrivial topology. For example, in two-dimensional quantum field theory on complex plane, the $n$-th R\'enyi entropy of two intervals requires the partition function on a Riemann surface of genus $(n-1)$, which is hard to compute\cite{Calabrese:2009ez}. Nevertheless, when two intervals are short and far apart, one can use the operator product expansion(OPE) of the twist operators to compute the partition function order by order in the cross ratio\cite{Headrick:2010zt,Calabrese:2010he,Rajabpour:2011pt,Chen:2013kpa}.
 Especially for the large $c$ CFT dual to the AdS$_3$ gravity, the vacuum module contribution dominates the partition  function in the large central charge limit. This leads to a lot of study on the two-interval R\'enyi entropy in the large $c$ CFTs, which sheds new light on the AdS$_3$/CFT$_2$ correspondence\cite{OPE}.

The large distance expansion of the R\'enyi entropy can be applied to the CFTs in higher dimensions. In \cite{Cardy:2013nua}, the mutual information of two disjoint spheres in the 3D and 4D massless scalar theory has been studied to the leading order in the large distance expansion.  The result is consistent with the lattice computation in \cite{Shiba:2012np}  and the study in \cite{Casini:2008wt}. In \cite{Agon:2015twa}, the computation in 3D case has been pushed to the next-to-leading order. For other study on the mutual information, see \cite{Agon:2015ftl,Maghrebi:2015,Schnitzer:2014zva,Herzog:2014fra}.

Furthermore, it was emphasized in \cite{Chen:2016mya, Long:2016vkg} that  the mutual information could be reorganized in terms of the conformal blocks
\be
I_{A,B}=\sum_{\{\D,J\}}b_{\D,J}G_{\D,J},
\ee
where the summation is over all the primary modules in the replicated theory. The key ingredient in the  study is the OPE of spherical twist operator\cite{Hung:2014npa,Long:2016vkg,Bianchi:2015liz} in terms of the primary operators in the replicated theory. This allows the systematic study of the mutual information. Different from 2D large $c$ CFT, one needs to consider the contribution from other modules, not just the vacuum module, in a general CFT. Even for the free CFTs of the free scalar discussed in \cite{Chen:2016mya}, the spectrum becomes quite involved in a replicated theory. One lesson is that there are much more primary operators in the replicated theory than the ones in original CFT. Even though the complete classification of the spectrum in the replicated theory is involved, the primary operators of the first few lowest dimensions can be constructed straightforwardly. These operators give the first few leading order contributions to the mutual information. For example, if the original CFT has conformal families $\{O^{(i)}\}$, then the leading contribution from each family to the mutual information is from the operators of bilinear form $O^{(i)}_{j_1}O^{(i)}_{j_2}$ with $j_1,j_2$ labeling the replica, and the next-leading order one could come from the bilinear operators with one derivative. It turns out that the coefficients before the conformal blocks from these operators are of universal forms, determined by the scaling dimensions and spacetime dimensions. Consequently such universal behavior persists in  the first few leading order contributions to the mutual information. 

The feasibility to read the universal behavior in the mutual information is due to the simplification in the $n\to 1$ limit of the R\'enyi mutual information. This issue has been discussed in \cite{Agon:2015ftl}, where it has been shown that the leading order mutual information from a scalar-type operator take a universal form. The analysis in \cite{Agon:2015ftl} relied on the subtle analytic continuation. In this work, we propose a $1/n$ prescription to simplify the discussion and apply it to more general cases, including the bilinear operators with derivatives, the bilinear operators made of the vector-type, or tensor-type or fermion-type operators, and even the operators of quartic form. This allows us to read more universal results beyond the leading order. For example, for the scalar-type operators, not only the leading order term, but also the sub-leading order terms coming from the bilinears with derivatives
are of universal forms. For the fermionic-type operators, the leading order contribution is of universal form as well.

The remaining parts of this paper are organized as follows. In the next section, after giving a brief review of spherical twist operator and its OPE expansion, we introduce the $1/n$ prescription in detail and discuss its application to the scalar-type, vector-type and tensor-type operators. In Section 3, we revisit the free scalar theory in three and four dimensions and compare our result with the improved numerical analysis. In Section 4, we study the mutual information in the free fermion theory. We end with conclusions and discussions in Section 5. In a few appendices, we collect some technical details, including our convention and improved treatment on numerical studies on free scalars and fermions. 

\section{Mutual information and conformal block}

 A twist operator is a co-dimension two operator and introduce the branch cut at the entangling surface in the path integral over the $n$-fold replicated theory.
Let us focus on the case that the entangling surface $A$ is a single sphere of  radius $R$ and discuss the spherical twist operator.  In order to compute $\Tr \r_A^n$, one can compute the partition function of the CFT in the $n$-fold replicated geometry. Equivalently  one may introduce a spherical twist operator $\cT_n$ and compute the correlation function of the twist operator in the $n$-fold replicated theory.
In two dimension, the twist operators are the local primary operator defined at the branch points. In higher dimensions, the twist operators are non-local surface operators.

The twist operator is a defect operator of co-dimension two. Just like nonlocal Wilson loop or surface operators\cite{Shifman:1980ui,Berenstein:1998ij,Gomis:2009xg,Chen:2007zzr}, the defect operator can be expanded in
terms of the local operators. In a conformal field theory, the local operators can be classified into conformal families, each one including a primary field and its descendants. For a defect operator\cite{Billo:2016cpy,Gadde:2016fbj} of co-dimension $q$ and  preserving $SO(d-q+1,1)\times SO(q-1,1)$ symmetry, it can be expanded into the sum of local operators\cite{Long:2016vkg}
\be
\cD =<\cD>\sum_{\{\D,J\}}c_{\D,J}(O_{\D,J}+\mbox{its descendants}).
\ee
where $<\cD>$ is the expectation value of the defect operator and gives the partition function of the defect CFT. The coefficients $c_{\D,J}$ are determined by the one-point function of the operator $O_{\D,J}$ in the presence of the defect operator. 

For a spherical twist operator,
it can be expanded in a similar way
\be
\cT_n=<\cT_n>\sum_{\{\D,J\}}c_{\D,J}Q[O_{\D,J}],
\ee
where $Q[O_{\D,J}]$ denotes all the operators generated from the primary operator $O_{\D,J}$ of dimension $\D$ and spin $J$.
Note that the summation is over all the primary operators in the $n$-replicated CFT. The coefficient $c_{\D,J}$ is read from
 the one-point function of the primary operator in the presence of the spherical twist operator. Equivalently it can be computed by the one-point function of the primary operator in the conical geometry. In general, this one-point function may not be easy to compute. 

In this work, we consider a CFT in its vacuum state   and  we let the entangling regions  be  spheres at a constant time slice. We consider  two disjoint spheres  such that the operator product expansion(OPE) of the twist operators can be applied.
Let the two spheres be
\be
A=\{t=0, \vec{x}^2\leq R^2\}, \hs{3ex}B=\{t=0, (\vec{x}-\vec{x}_0)^2\leq R'^2\}.
\ee
For simplicity, we use conformal symmetry to set $R=R'$ and $\vec{x}_0=(1,0,\cdots,0)$. The only independent conformal invariant quantity is the cross ratio
\be
z=\bar{z}=4R^2,\hs{5ex}u=z\bar{z},\hs{3ex}v=(1-z)(1-\bar{z}).
\ee
In the disjoint case, we have $0<z<1$. More generally, if the radii of two spheres are different, and their distance is $r$, then the cross ratio is
\be
z=\bar{z}=\frac{4RR'}{r^2-(R-R')^2}.
\ee

We are interested in the mutual information of two spheres. It can be read from the $n\to 1$ limit of the R\'enyi mutual information, which is defined to be
\be
I^{(n)}_{A,B}= S^{(n)}_{A}+S^{(n)}_{B}-S^{(n)}_{A\cup B}.
\ee
The R\'enyi entropy of $A\cup B$ is
\be
S^{(n)}_{A\cup B}=\frac{1}{1-n}\log \la \cT_n(A\cup B)\ra.
\ee
When the two regions are far apart, we may evaluate the expectation value of the twist operator approximately by
\bea
\frac{ \la \cT_n(A\cup B)\ra}{ \la \cT_n(A)\ra   \la \cT_n(B)\ra} =\frac{ \la \cT_n(A) \cT_n(B)\ra}{ \la \cT_n(A)\ra   \la \cT_n(B)\ra}&=& \sum_{\{\D,J\}}c^2_{\D,J}\la Q[O_{\D,J}](A)Q[O_{\D,J}](B) \ra\nn\\
 &=&\sum_{\{\D,J\}}s_{\D,J}G_{\D,J}(u,v).
\eea
where the building block is the two-point function of the primary module and is related to the conformal block\cite{Dolan:0011,Dolan:0309}. The coefficient $s_{\D,J}$ is given by
\be
s_{\Delta,J}=f_{\Delta,J}\sum_{\mathcal{O}_{\Delta,J}}\frac{a^2_{\Delta,J}}{N_{\Delta,J}}, \label{sDJ}
\ee
where the summation is over all the primary operators with the same $(\D,J)$ in the replicated theory, $a_{\D,J}$ is determined by the one-point function of the operator $O_{\D,J}$ in the planar conical geometry
\be
\la O_{\D,J}(x)\ra_n=a_{\D,J}\frac{T_J}{|x|^\D}
\ee
with $T_J$ being a kind of tensor structure,
and $N_{\D,J}$ is the normalization factor in the two-point function in the flat spacetime
\be
\langle{O}_{\Delta,J}(x){O}_{\Delta,J}(x^{\prime})\rangle=N_{\Delta,J}\frac{T^{\prime}_{J}(x-x^{\prime})}{(x-x^{\prime})^{2\Delta}}
\ee
with $T'_J$ being the tensor structure relating to the operator with spin $J$.  The explicit form of $T_J$ and $T'_J$ can be found in Appendix A. The remaining coefficient $f_{\D,J}$ could be determined by considering one spherical operator and mapping it to a half plane. It depends only on the tensor structure of the operator. The details to fix these coefficients can be found in Appendix B. More explicitly, we find that\footnote{Originally, these coefficients are fixed in  Lorentz CFTs\cite{Long:2016vkg}. However, we will work in Euclidean CFTs so we have flipped the sign of the odd spin coefficients due to the Wick rotation.}
\be
s_{\D,0}=\sum_{O_{\D,0}}\frac{a_{\D,0}^2}{N_{\D,0}},\hs{3ex}
s_{\D,1}=\sum_{O_{\D,1}}\frac{a_{\D,1}^2}{N_{\D,1}},\hs{3ex}
s_{\D,2}=d(d-1)\sum_{O_{\D,2}}\frac{a_{\D,2}^2}{N_{\D,2}},
\ee
\bea
s_{\Delta,3}&=&(d-1) (d+2)\sum \frac{a_{\Delta,3}^2}{N_{\Delta,3}},\\
s_{\Delta,4}&=&(d-1) (d+1) (d+2) (d+4)\sum \frac{a_{\Delta,4}^2}{N_{\Delta,4}}.
\eea

In terms of the conformal blocks, the R\'enyi mutual information can be expressed by \be
I^{(n)}_{A,B}=-\frac{1}{1-n}\log (1+\sum_{\{\D,J\}}s_{\D,J}G_{\D,J}),
\ee
and the mutual information is just
\be
I_{A,B}=\sum_{\{\D,J\}}b_{\D,J}G_{\D,J}(u,v) ,
\ee
with the coefficient $b_{\D,J}$ being related to the expansion of $s_{\D,J}$ with respect to $(n-1)$
\be
b_{\D,J}=s'_{\D,J}(n=1).
\ee
This is the conformal block expansion of the mutual information. The coefficients $b_{\D,J}$ encode the dynamical information of corresponding conformal field theory.
As the conformal block in the diagonal limit is approximated by\cite{Matthijs:1305}
\be
G_{\D,J}(z) \simeq z^\D+ \cdots,
\ee
the leading contribution to the mutual information is from the primary operator with the lowest dimension and nonvanishing coefficient.

In a CFT, the primary module is characterized by the conformal dimension $\D$ and the spin $J$. Moreover, as we are going to consider the replicated theory, the spectrum becomes more complicated.   Besides the primary operators in the original CFT, there are many more primary operators in the replicated theory. The spectrum of the replicated theory could be constructed from the conformal families of original CFT, taking into account of the fusion. To explain the possible primary operators in the replicated theory, let us simply assume
all the primary operators in the original CFT to be the scalar type, and label them according to their scaling dimensions $O^{(1)},O^{(2)},\cdots O^{(m)}$ with $\D_1\leq \D_2 \leq ...\D_m$. From each family, one may construct the primary operators in the replicated theory, similar to the discussion on free scalar in \cite{Chen:2016mya}. However, there are primary operators from different conformal families. The simplest primary operator is the ones $O^{(i)}_j$, where $j$ labels the replica. This kind of operator has no contribution to the mutual information. The simplest primary operator with nonvanishing contribution is the one of bilinear form from the same conformal family: $O_{j_1}^{(i)}O_{j_2}^{(i)}$ where $j_1\neq j_2$ represent the replicas. There do exist the bilinear primary operators made of different conformal families: $O_{j_1}^{(i_1)}O_{j_2}^{(i_2)} (i_1\neq i_2)$, but they have no contribution to the mutual information. Depending on the conformal dimension, the next-to-leading order contribution may come from a vector
\be
J^{(i)}_{\m,j_1j_2}=O^{(i)}_{j_1}\p_\m O^{(i)}_{j_2}-(j_1 \leftrightarrow j_2)
\ee
or the operator of cubic form if there exists fusion among conformal families
\be
C_{j_1j_2j_3}=O^{(i_1)}_{j_1}O^{(i_2)}_{j_2}O^{(i_3)}_{j_3}, 
\ee
or the primary operator of quartic form
\be
Q_{j_1j_2j_3j_4}=O^{(i_1)}_{j_1}O^{(i_2)}_{j_2}O^{(i_3)}_{j_3}O^{(i_4)}_{j_4}.
\ee
In the last two cases, the operators in the construction may from the same conformal family or the different one,
  and the indices $j_1,j_2,\cdots$ can be equal but the nonvanishing contribution  is from the operators at different replicas. One may construct more involved primary operators by including the derivatives.  In any case, we can still classify the primary operators in the replicated theory by their conformal dimensions and spins under the  conformal group.



From the above discussion, it is evident that the leading contribution comes from the  primary module $\{O^{(1)}\}$ with the lowest conformal dimension, and the next-to-leading ones come either from the bilinear operator of another conformal family $\{O^{(2)}\}$, or the bilinear one with derivatives or the quartic operators\footnote{In this work, we do not consider the fusion of the primary modules, which may lead to next-to-leading order contributions.}. In any case, we can focus on the discussion of a single conformal family, keeping in mind the one with the lowest dimension.

\subsection{$1/n$ prescription}

With a primary module at hand, we need to compute its coefficient $b_{\D,J}$ before the conformal block. Naively we need to read the coefficient $s_{\D,J}$ given in (\ref{sDJ}) first. It  is generically very hard to compute, as it requires the exact information of the one-point function of a primary operator in the conical geometry. However, if we are only interested in the mutual information, we can expand the coefficient in the orders of $(n-1)$. It turns out the discussion can be simplified and the coefficients $a_{\D,J}$ and $s_{\D,J}$ under the $n\to 1$ limit are related to the correlation functions in the flat spacetime. This fact allows us to read the coefficient $b_{\D,J}$ for a primary module in the replicated theory.

Let $G_n$ be any periodic function in the conical geometry $\cC_n$, whose angular direction is identified as $\th \simeq \th + 2\pi n$. Obviously the function satisfies:
\be
G_n(r,\th, y^i)=G_n(r,\th+2\pi n, y^i),\ee
and its $n\to 1$ limit gives the usual function on the flat space
\be
G_1(r,\th,y^i)=\lim_{n\to 1}G_n(r,\th, y^i). \label{nto1}
\ee
We can do Fourier expansion on $G_n,G_1$ along $\th$ direction respectively
\bea
G_n(r,\th,y^i)&=&\sum_{k\in Z}g_n(r,k,y^i)e^{ik\th/n},\\
G_1(r,\th,y^i)&=&\sum_{k\in Z}g_1(r,k,y^i)e^{ik\th}.
\eea
Due to the relation (\ref{nto1}), we find that
\be
g_n(k)=g_1(k)+(n-1)f(k)+\cO((n-1)^2),
\ee
where  $f(k)$ is independent of $n$ and is not divergent.
 This leads us to
\bea
G_n(\th)&=&G_1(\theta/n)+(n-1)\sum_{k\in Z}f(k)\exp\{ik\theta/n\}+\cO((n-1)^2).
\eea
 We call this relation $1/n$ prescription. It tells us that the periodic function in the conical geometry in the $n\to 1$ limit is related to the function in the flat space by dividing the angular variable by $n$. This relation may help us to read the coefficient $a_{\D,J}$ in the limit of $n\to1$ directly.

 Furthermore, in order to determine the coefficient $b_{\D,J}$, we need to consider the following summation
 \be
\lim_{n\to1}\sum_{q=1}^{n-1} \frac{(G_n(2\pi q))^2}{n-1}.
\ee
The summation actually satisfies
\be
\lim_{n\to1}\sum_{q=1}^{n-1} \frac{(G_n(2\pi q))^2}{n-1}=\lim_{n\to1}\left(\sum_{q=1}^{n-1}\frac{(G_1(2\pi q/n))^2}{n-1}+ 2\sum_{k\in Z}h(k)(\sum_{q=1}^{n-1}\exp(2\pi iqk/n)) +\cO(n-1)\right). \label{sum}
\ee
where
\be
h(k)=\sum_{w\in Z}f(k-w)g_1(w).
\ee
In (\ref{sum}), the summation over $q$ actually gives
\be
\sum_{q=1}^{n-1}\exp\{2\pi ikq/n\}=\left\{\begin{array}{ll}
n-1,& k=jn, \hs{3ex}j\in Z\\
-1, & \mbox{otherwise}
\end{array}\right.
\ee
Then  the second term in the summation can be recast into
\be
2\sum_{w\in Z}\sum_{k\in Z}(nf(nk-w)g_1(w)-f(k-w)g_1(w)).
\ee
 In the  $n\to 1$ limit, this term is vanishing. Consequently, we need only to consider the first term in the summation (\ref{sum}) in the $n\to 1$ limit
  \begin{equation}
\lim_{n\rightarrow 1}\sum_{q=1}^{n-1}\frac{G_n^2(2\pi q)}{n-1}=\lim_{n\rightarrow 1}\sum_{q=1}^{n-1}\frac{G_1^2(2\pi q/n)}{n-1}
\end{equation}

 Let us assume the primary operator of the lowest conformal dimension in the original theory to be $O$. Its one-point function in the conifold geometry is vanishing in the $n\to 1$ limit so that it does not contribute to the mutual information directly. However, the operator $O_{j_1} O_{j_2}$ is primary in the replicated theory and its one-point function in the conifold geometry is not vanishing. In this case, we have to compute the two-point function of the operator $O$ in the conical geometry. In general, such a two-point function is
quite complicated. It cannot be determined by the symmetry, but instead depends on the dynamics of the theory.  However, if we just interested in the mutual information, we need only to extract the information in the $n \to 1$ limit. In other words, we may extract the information which is not vanishing in the limit. Such information is nicely encoded in the two-point function of the primary operator in the flat space by using the above $1/n$ prescription. For more general primary operators in the replicated theory, we can have similar discussion.

In the next few subsections, we present the coefficients of bilinear operators constructed from a single conformal family of scalar-type, vector-type and tensor-type operators respectively. However note that the $1/n$ prescription can be applied to more general types of primary operators as it is actually based on the $(n-1)$ expansion.

\subsection{Scalar type operator}

Let us first consider a scalar-type primary operator $O$ of the lowest dimension $\Delta$. Its  two-point function in flat spacetime is
\be
<O(x)O(x')>=\frac{N_{\Delta,0}}{|x-x'|^{2\Delta}},
\ee
then the two-point function on the conifold geometry is
\be
<O(x)O(x')>_n=\frac{N_{\Delta,0}}{|x-x'|_n^{2\Delta}}+\mathcal{O}(n-1)
\ee
where $|x-x'|_n^{2}$ is related to $|x-x'|^2$ by replacing $\theta-\theta'$ with  $\frac{\theta-\theta'}{n}$ according to $1/n$ prescription.
Just as a free scalar, the operator  giving the leading order contribution to the mutual information is of the type $O^{(s)}_{j_1j_2}=O_{j_1}O_{j_2}$, where $j_1,j_2$ label the replicas of the operators. The one-point function of $O^{(s)}_{j_1j_2}$ is
\be
<O^{(s)}_{j_1j_2}>_n=\frac{N_{\Delta,0}}{(4r^2s^2_{j_1j_2})^{\Delta}}+\mathcal{O}(n-1)
\ee
where we have defined the coordinates
\be
x=(r,\theta,\vec{y}), \hs{3ex}\th_{ij}=\th_i-\th_j,
\ee
and
\be
s_{j_1j_2}\equiv \sin\frac{\pi(j_1-j_2)}{n}.
\ee
Hence we read its one-point function coefficient
\be
a^{(s)}_{2\Delta,0}=\frac{N_{\Delta,0}}{(4s^2_{j_1j_2})^{\Delta}}+\mathcal{O}(n-1),
\ee
with the normalization coefficient being
\be
N^{(s)}_{2\Delta,0}=N_{\Delta,0}^2
\ee
Then we find the function
\be
s^{(s)}_{2\Delta,0}=\frac{n}{2}\sum_{j=1}^{n-1}\frac{1}{(4s^2_{j_1j_2})^{2\Delta}}+\mathcal{O}((n-1)^2)
\ee
where $\th_j=2\pi j$. 
Using the integral representation given in Appendix A.4,  it is straightforward to get
\be
b^{(s)}_{2\Delta,0}=\lim_{n\to1}\frac{1}{n-1}s^{(s)}_{2\Delta,0}=\frac{\sqrt{\pi}\Gamma[2\Delta+1]}{4^{2\Delta+1}\Gamma[2\Delta+\frac{3}{2}]}.\label{b2d0}
\ee
This is the same as the one in \cite{Agon:2015ftl}.

For the same kind of operator $O$, we may construct a primary operator with spin 1 by\footnote{To simplify the notation, we omit the indices $j_1,j_2$ labelling the replicas in the constructed operators in this section.}
\be
J^{(s)}_{\mu}=O_{j_1}\partial_{\mu}O_{j_2}-(j_1\leftrightarrow j_2).
\ee
It has conformal dimension $2\Delta+1$ and normalization $4\Delta$. Its one-point function in the planar conifold geometry is
\be
<J^{(s)}_{a}>_n=-\frac{1}{n}2^{1-2\Delta}\Delta\cos\frac{\theta_{j_1j_2}}{2n}\csc^{1+2\Delta}\frac{\theta_{j_1j_2}}{2n}\frac{\epsilon_{ab}n^b}{|x|^{2\Delta+1}}+\mathcal{O}(n-1).
\ee
In a similar way we may read the coefficient before the spin 1 conformal block
\be
b^{(s)}_{2\Delta+1,1}=-\frac{\sqrt{\pi}\Delta\Gamma[1+2\Delta]}{2^{3+4\Delta}\Gamma[2\Delta+\frac{5}{2}]}.\label{b2dp11}
\ee

There is a spin 2 operator which contributes to the mutual information as well
\be
T^{(s)}_{\mu\nu}=\frac{1}{2}P^{\alpha\beta}_{\mu\nu}(\partial_{\alpha}O_{j_1}\partial_{\beta}O_{j_2}
-\frac{\Delta}{\Delta+1}O_{j_1}\partial_{\alpha}\partial_{\beta}O_{j_2}+(j_1\leftrightarrow j_2)),
\ee
where $P^{\alpha\beta}_{\mu\nu}$ is the projector\footnote{The explicit forms of the projectors used in this subsection can be found in Appendix A.1.} to project the tensor field to its symmetric traceless part.
It has dimension $2\Delta+2$ and normalization $\frac{4\Delta^2(1+2\Delta)}{1+\Delta}$. Its one-point function in the planar conifold geometry is
\be
<T^{(s)}_{ij}>_n=\frac{\Delta(1+n^2+2\Delta)(1+\Delta+\Delta c_{j_1j_2} )}{2^{1+2\Delta}dn^2(1+\Delta)s^{2+2\Delta}_{j_1j_2}}\frac{\delta_{ij}}{|x|^{2\D+2}}+\mathcal{O}(n-1),
\ee
where
\be
c_{j_1j_2}\equiv \cos\frac{\theta_{j_1j_2}}{n}.
\ee
After some algebra, we find
\be
b^{(s)}_{2\Delta+2,2}=\frac{(d-1)\sqrt{\pi}(2+4\Delta+3\Delta^2)\Gamma[2\Delta+3]}{d(2\Delta+1)^22^{4\Delta+5}\Gamma[2\Delta+\frac{7}{2}]}.\label{b2dp22}
\ee

The spin 3 operator made of two $O$'s could be of the form
\be
O^{(s)}_{\mu_1\mu_2\mu_3}=P^{\alpha_1\alpha_2\alpha_3}_{\mu_1\mu_2\mu_3}(\partial_{\alpha_1\alpha_2\alpha_3}O_{j_1}O_{j_2}
-\frac{\Delta+2}{\Delta}\partial_{(\alpha_1\alpha_2}O_{j_1}\partial_{\alpha_3)}O_{j_2}-{j_1\leftrightarrow j_2}).
\ee
It has dimension $2\Delta+3$ and normalization $192(\Delta+1)(\Delta+2)(2\Delta+3)$. The coefficient from the one-point function in the planar conifold geometry is
\bea
a^{(s)}_{2\Delta+3,3}&=&\frac{2^{1-2\Delta}c_{j_1j_2}}{(d+2)n^3s_{j_1j_2}^{2\Delta+3}}((3 + 2 \Delta) (1 + \Delta(3 + \Delta)) +
 n^2 (9 + \Delta (11 + 3 \Delta))\nn\\&&+\Delta (\Delta(3 + 2 \Delta) + n^2 (4 + 3 \Delta))(1-2s_{j_1j_2}^2))+\mathcal{O}(n-1),
\eea
from which we find
\be
b^{(s)}_{2\Delta+3,3}=\frac{-(d-1)\sqrt{\pi}((1 + \Delta) (2 + \Delta) (6 +12 \Delta + 5 \Delta^2)) \Gamma[2\Delta+1]}{(d+2)2^{4\Delta+5}(3 + 2 \Delta) \Gamma[  2 \Delta+\frac{9}{2}]}.
\ee
For a general spin $s$ operator made of two $O$'s could be of the form
\be
O^{(s)}_{\mu_1\cdots\mu_s}=P^{\alpha_1\cdots\alpha_s}_{\mu_1\cdots\mu_s}\sum_{k=0}^s(-1)^k\frac{\Gamma[\Delta+s]\Gamma[\Delta]}{\Gamma[\Delta+s-k]\Gamma[\Delta+k]}
\partial_{(\mu_1\cdots\mu_{s-k}}O_{j_1}\partial_{\mu_{s-k+1}\cdots\mu_s)}O_{j_2}.
\ee
It has dimension $2\Delta+s$ and  normalization $\frac{2^s \Gamma[1 + s] \Gamma[-1 + 2 s + 2 \Delta]}{\Gamma[-1 + s + 2 \Delta]}$.
For a spin 4 operator, we find
\bea
a^{(s)}_{2\Delta+4,4}&=&\frac{1}{(2 + d) (4 + d) n^4 4^{\Delta}(1 + \Delta)s_{j_1j_2}^{2\Delta}}(4 \Delta (3 n^4 (2 + \Delta)^2 + \Delta^2 (3 + 2 \Delta) (5 +
      2 \Delta) \nn\\&&+
   2 n^2 (3 + 2 \Delta) (2 + \Delta (8 + 3 \Delta)))-2 (3 + 2 \Delta) ((1 + 2 \Delta) (5 + 2 \Delta) (1 +
       2 \Delta (1 + \Delta)) \nn\\&&+
    3 n^4 (3 + 2 \Delta (3 + \Delta)) +
    n^2 (34 +
       4 \Delta (24 + \Delta (23 +
             6 \Delta)))) \csc^2s_{j_1j_2} \nn\\&&+(1+\Delta)(3 + 2 \Delta) (5 + 2 \Delta) (3 + 3 n^4 +
   4 \Delta (2 + \Delta) + 6 n^2 (3 + 2 \Delta))\csc^4s_{j_1j_2})+\mathcal{O}(n-1).\nn
\eea
and
\be
b^{(s)}_{2\Delta+4,4}=\frac{\sqrt{\pi}(d^2-1) (1 + \Delta) (2 + \Delta) (3 + \Delta)(72 + 204 \Delta + 155 \Delta^2 + 35 \Delta^3)\Gamma[1 + 2 \Delta]}{2^{4\Delta+7}(2 + d) (4 + d) (3 + 2 \Delta) (5 + 2 \Delta) \Gamma[
  \frac{11}{2} + 2 \Delta]}.
\ee

\subsection{Vector-type operator}

For the vector-type operator, it has spin 1 and its two-point function in flat space is
\be
<J_{\mu}(x)J_{\nu}(x')>=N_{\Delta,1}\frac{I_{\mu\nu}(x-x')}{|x-x'|^{2\Delta}}
\ee
where
\be
I_{\mu\nu}(x)=\delta_{\mu\nu}-2n_\mu n_\nu, \hs{3ex}\mbox{with}\hs{2ex}n_\mu=\frac{x_\mu}{|x|}.
\ee
Note that from the spin 1 operator $J_{\mu}$, we can construct three kinds of primary operators due to the following decomposition
\be
\Yvcentermath1
\yng(1)\otimes\yng(1)=\yng(2)\oplus\bullet\oplus\yng(1,1) .
\ee
The three kinds of operators are of the following forms respectively
\be
O^{(v)}_{\mu\nu}=P^{\alpha\beta}_{\mu\nu}(J_{\alpha}^{j_1}J_{\beta}^{j_2}),\hs{3ex} O^{(v)}=J_{\mu}^{j_1}J_{j_2}^{\mu}, \hs{3ex} O'^{(v)}_{\mu\nu}=J_{\mu}^{j_1}J_{\nu}^{j_2}-(\mu\leftrightarrow\nu).
\ee
It turns out that the third kind operator does not contribute to the mutual information.
After similar analysis as the scalar operator, we find
\bea
N^{(v)}_{2\Delta,0}=N_{\Delta,1}^2 d,& &a^{(v)}_{2\Delta,0}=(d-2)\frac{N_{\Delta,1}}{(4s^2_{j_1j_2})^{\Delta}}+\mathcal{O}(n-1),\nn\\
 N^{(v)}_{2\Delta,2}=N_{\Delta,1}^2 ,
&& a^{(v)}_{2\Delta,2}=\frac{2}{d}\frac{N_{\Delta,1}}{(4s^2_{j_1j_2})^{\Delta}}+\mathcal{O}(n-1).\nn
\eea
Then we find the coefficients of two different kinds of the operators before the conformal block are respectively
\bea
b^{(v)}_{2\Delta,0}&=&\frac{(d-2)^2}{d}\frac{\sqrt{\pi}\Gamma[2\Delta+1]}{4^{2\Delta+1}\Gamma[2\Delta+\frac{3}{2}]},\\ b^{(v)}_{2\Delta,2}&=&\frac{(d-1)}{d}\frac{\sqrt{\pi}\Gamma[2\Delta+1]}{4^{2\Delta}\Gamma[2\Delta+\frac{3}{2}]}.
\eea

\subsection{Spin 2 case}

In this case, the typical example is the stress tensor. The two-point function of such operator is of the form
\be
<T_{\mu\nu}(x)T_{\rho\sigma}(x')>=N_{\Delta,2}\frac{I_{\mu\nu,\rho\sigma}(x-x')}{|x-x'|^{2\Delta}},
\ee
where
\be
I_{\mu\nu,\rho\sigma}(x)=
\frac{1}{2}(I_{\mu\rho}(x)I_{\nu\sigma}(x)+I_{\mu\sigma}(x)I_{\nu\rho}(x))-\frac{1}{d}\delta_{\mu\nu}\delta_{\rho\sigma}.
\ee
From the simple product of two spin-2 operators, we can construct six kinds of operators due to the decomposition
\be
\Yvcentermath1
\yng(2)\otimes\yng(2)=\yng(4)\oplus\yng(2)\oplus\bullet\oplus\yng(3,1)\oplus\yng(2,2)\oplus\yng(1,1)
\ee
Only the first three types of the operators contribute to the mutual information.
They are of the following forms respectively
\bea
&&O^{(t)}=T_{\mu\nu}^{j_1}T^{\mu\nu}_{j_2},\nn\\ &&O^{(t)}_{\mu\nu}=P^{\alpha\beta}_{\mu\nu}(T_{\alpha\gamma}^{j_1}T_{\beta}^{\gamma j_2}),\nn\\
&&O^{(t)}_{\mu\nu\rho\sigma}=P^{\alpha\beta\gamma\delta}_{\mu\nu\rho\sigma}(T_{\alpha\gamma}^{j_1}T_{\beta\delta}^{j_2}).
\eea
In a general dimension, we find
\bea
N^{(t)}_{2\Delta,4}=N_{\Delta,2}^2, &&a^{(t)}_{2\Delta,4}=\frac{4N_{\Delta,2}}{(d+2)(d+4)(4s_{j_1j_2}^2)^{\Delta}},\nn\\ N^{(t)}_{2\Delta,2}=\frac{(d+4)(d-2)}{4d}N_{\Delta,2}^2,&&a^{(t)}_{2\Delta,2}=\frac{(d-2)N_{\Delta,2}}{d}\frac{1}{(4s_{j_1j_2}^2)^{\Delta}},\nn\\ N^{(t)}_{2\Delta,0}=\frac{(d+2)(d-1)}{2}N_{\Delta,2}^2,&&a^{(t)}_{2\Delta,0}=\frac{(d-2)(d-1)N_{\Delta,2}}{2}\frac{1}{(4s_{j_1j_2}^2)^{\Delta}},
\eea
and their coefficients before the conformal block
\bea
b^{(t)}_{2\Delta,4}&=&\frac{\sqrt{\pi}(d^2-1)\Gamma[2\Delta+1]}{(8 + 6 d + d^2)2^{4\Delta-2}\Gamma[\frac{3}{2} + 2 \Delta]},\nn\\
b^{(t)}_{2\Delta,2}&=&\frac{\sqrt{\pi}(d-2)(d-1)\Gamma[2\Delta+1]}{(d+4)2^{4\Delta}\Gamma[\frac{3}{2} + 2 \Delta]},\label{spin2}\\
b^{(t)}_{2\Delta,0}&=&\frac{\sqrt{\pi}(d-2)^2(d-1)\Gamma[2\Delta+1]}{(d+2)2^{4\Delta+3}\Gamma[\frac{3}{2} + 2 \Delta]}.\nn
\eea

\section{Free scalar theory}

The above discussion on the $1/n$ prescription is quite general and could be applied to any CFT. Once we are clear of the spectrum of the CFT, we can read the leading order contribution to the mutual information when the OPE of the defect operator is applicable. To illustrate the picture, let us first consider the generalized free theory, whose spectrum includes only a scalar-type operator $O$ of dimension $\D$. 
In this generalized free theory,  the primary operator $O$ has the correlation functions similar to the free scalar theory:
\bea
<O(x_1)O(x_2)>&=&G(x_{12})=\frac{1}{|x_1-x_2|^{2\Delta}},\nn\\
<O(x_1)O(x_2)O(x_3)>&=&0,\\
<O(x_1)O(x_2)O(x_3)O(x_4)>&=&G(x_{12})G(x_{34})+G(x_{13})G(x_{24})+G(x_{14})G(x_{23}).\nn
\eea
For a replicated theory, there are many kinds of primary operators. Among the ones without considering the derivatives,  the operators of dimension $2\Delta$ and nonvanishing one-point functions in the conical geometry include two types
\be
O^2_j(x),\hs{2ex} O_{j_1}O_{j_2}, \hs{2ex}(j_1\neq j_2).
\ee
 All the cubic operators of dimension $3\Delta$  have vanishing one-point functions, and the quartic operators of dimension $4\Delta$ and nonvanishing one-point functions in the conical geometry include five types
 \bea
 O^4_j, &O^3_{j_1}O_{j_2}, &O^2_{j_1}O^2_{j_2}\nn\\
 O^2_{j_1}O_{j_2}O_{j_3}& O_{j_1}O_{j_2}O_{j_3}O_{j_4}& (j_i \neq j_k, \hs{2ex} \mbox{if}\hs{2ex}i\neq k)
 \eea
 Their one-point functions can be computed following the $1/n$ prescription
\bea
<O^2_j(x)>_n&=&\mathcal{O}(n-1)\nn\\
<O_{j_1}O_{j_2}>_n&=&\frac{1}{(4r^2s_{j_1j_2}^2)^{\Delta}}+\mathcal{O}(n-1)\nn\\
<O^4_j>_n&=&\mathcal{O}(n-1)\nn\\
<O^3_{j_1}O_{j_2}>_n&=&\mathcal{O}(n-1)\\
<O^2_{j_1}O^2_{j_2}>_n&=&\frac{2^{1-4\Delta}}{r^{4\Delta}s_{j_1j_2}^{4\Delta}}+\mathcal{O}(n-1)\nn\\
<O^2_{j_1}O_{j_2}O_{j_3}>_n&=&\frac{2^{1-4\Delta}}{r^{4\Delta}s_{j_1j_2}^{2\Delta}s_{j_1j_3}^{2\Delta}}+\mathcal{O}(n-1)\nn\\
<O_{j_1}O_{j_2}O_{j_3}O_{j_4}>_n&=&\frac{2^{-4\Delta}}{r^{4\Delta}}(\frac{1}{(s_{j_1j_2}s_{j_3j_4})^{2\Delta}}+\frac{1}{(s_{j_1j_3}s_{j_2j_4})^{2\Delta}}+\frac{1}{(s_{j_1j_4}s_{j_2j_3})^{2\Delta}})+\mathcal{O}(n-1)\nn
\eea
The ones with the one-point functions being simply proportional to $(n-1)$ give no contribution to the mutual information. The leading contribution to the mutual information comes from the bilinear operator $O_{j_1}O_{j_2}, \hs{2ex}(j_1\neq j_2)$ and the next-to-leading one comes from three kinds of quartic operators. Note that there could be contribution from the bilinear operators with a derivative, i.e. spin-1 operator in Section 2.2, whose contribution could be even at the same order or larger than the quartic operators.

Let us consider three dimensional free scalar theory. In this case the conformal dimension of the real scalar is  $\Delta=\frac{1}{2}$.
The leading order contribution to the mutual information is from the operator of bilinear  form $O_{j_1}O_{j_2}$, which has the coefficient
\be
b^{(3d)}_{1,0}=\frac{1}{12}.
 \ee
 The next-to-leading one come from three kinds of operators of quartic form, all having dimension two. The first kind is $O^2_{j_1}O^2_{j_2}$ with the coefficient $\frac{1}{60}$  which is same as (4.8) in \cite{Agon:2015twa}. The second kind is of the form $O^2_{j_1}O_{j_2}O_{j_3}$. As
\be
\frac{1}{2}\sum_{j_1\not=j_2\not=j_3,j_2>j_3}\frac{1}{4}\frac{1}{s_{j_1j_2}^2s_{j_1j_3}^2}+\mathcal{O}(n-1)^2=-\frac{1}{30}(n-1)+\mathcal{O}(n-1)^2,
\ee
this kind of operator gives a coefficient $-\frac{1}{30}$, which is three times  of (4.14) in \cite{Agon:2015twa}. The contribution of the third kind operator $O_{j_1}O_{j_2}O_{j_3}O_{j_4}$ can be found by noticing
\bea
\lefteqn{\sum_{j_1>j_2>j_3>j_4}\frac{1}{16}(\frac{1}{s_{j_1j_2}s_{j_3j_4}}+\frac{1}{s_{j_1j_3}s_{j_2j_4}}
+\frac{1}{s_{j_1j_4}s_{j_2j_3}})^2+\mathcal{O}(n-1)}\nn\\
&=&\frac{1}{8}\sum_{j_1>j_2>j_3>j_4}(\frac{1}{s_{j_1j_2}s_{j_3j_4}s_{j_1j_3}s_{j_2j_4}}+\frac{1}{s_{j_1j_2}s_{j_3j_4}s_{j_1j_4}s_{j_2j_3}}
+\frac{1}{s_{j_1j_3}s_{j_2j_4}s_{j_1j_4}s_{j_2j_3}})\nn\\
& &+\frac{1}{120}(n-1)+\mathcal{O}(n-1)^2.
\eea
It leads to (4.19) in \cite{Agon:2015twa}. The summation of all three kinds of quartic type operators gives the coefficient
\be
b^{(3d)}_{2,0}=\frac{1}{360}+\frac{1}{12\pi^2}.
 \ee
 For the three-dimensional free scalar, the spin-1 operator is of dimension two and gives the same order of contribution as the quartic operators. It has the coefficient
 \be
 b^{(3d)}_{2,1}=-\frac{1}{120}.
 \ee
To summarize, to the order $z^2$ the mutual information of three dimensional free scalar should be
\be
I_{A,B}=\frac{1}{12}G^{d=3}_{1,0}(z)-\frac{1}{120}G^{d=3}_{2,1}(z)+(\frac{1}{360}+\frac{1}{12\pi^2})G^{d=3}_{2,0}(z)+\cdots\label{3dscalar}
\ee
The first two conformal blocks are of the following forms
\bea
G^{d=3}_{1,0}(z)&=&\frac{z}{\sqrt{1-z}},\nn\\
G^{d=3}_{2,1}(z)&=&4(-2+\frac{2-z}{\sqrt{1-z}}).\nn
\eea

For the free scalars in other dimensions, it is straightforward to read the first few leading terms following the above general discussion. For the free scalar in four dimensions, the mutual information is\cite{Chen:2016mya}
\bea
I_{A,B}^{d=4}&=&\frac{1}{60}G^{d=4}_{2,0}-\frac{1}{420}G^{d=4}_{3,1}+\frac{1}{840}G^{d=4}_{4,2}+\frac{1}{840}G^{d=4}_{4,0}+\cdots\nn\\&=&\frac{z (-20 + (22 - 7 z) z) -
 2 (-1 + z) \log(1 - z) (-10 + 6 z + z \log(1 - z))}{140 (-1 + z) z}+\cdots\nn\\
 \label{4dscalar}
\eea
where the conformal blocks can be written in terms of generalized hypergeometric function.
Remarkably the first three coefficients before the conformal blocks are exactly the ones predicted in (\ref{b2d0}),(\ref{b2dp11}) and (\ref{b2dp22}).

The numerical study of the mutual information has been carried in \cite{Shiba:2012np}. We carry the numerical computation in higher precision in order to compare with the analytic result in the next-to-leading order. Some details on the numerical computation can be found in Appendix \ref{numerical}. In Fig. \ref{mutual3d} we draw the mutual information of two disks in three dimensional free scalar theory. We choose the cutoff $a=0.05$, the radius of the sphere $A$ and $B$ are $R=0.2$. One can change the distance $r$ to adjust the cross ratio $z$.  The points are from the lattice computation. The light blue curve is the theoretical result up to the leading order $z$. The pink line is theoretical result up to the next-to-leading order $z^2$. Note the next-to-leading order terms contribute significantly for $z$ is not too small ($z>0.12$).  

\begin{figure}
\centering
 \includegraphics[width=0.5\textwidth]{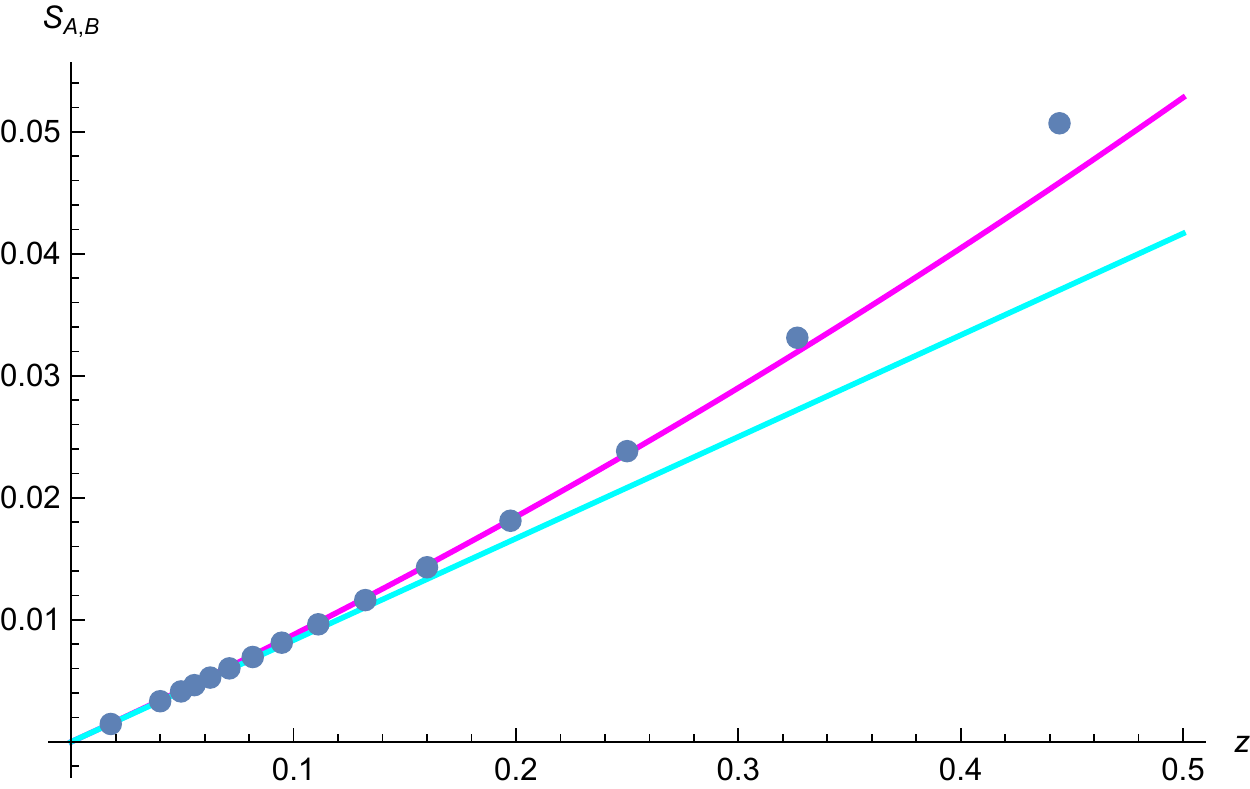}
     \caption{Mutual information of two disks for $3d$ free boson. The light blue curve is the theoretical result up to the leading order $z$. The pink line is theoretical result up to the next-to-leading order $z^2$. }\label{mutual3d}
\end{figure}

\begin{figure}
\centering
 \includegraphics[width=0.5\textwidth]{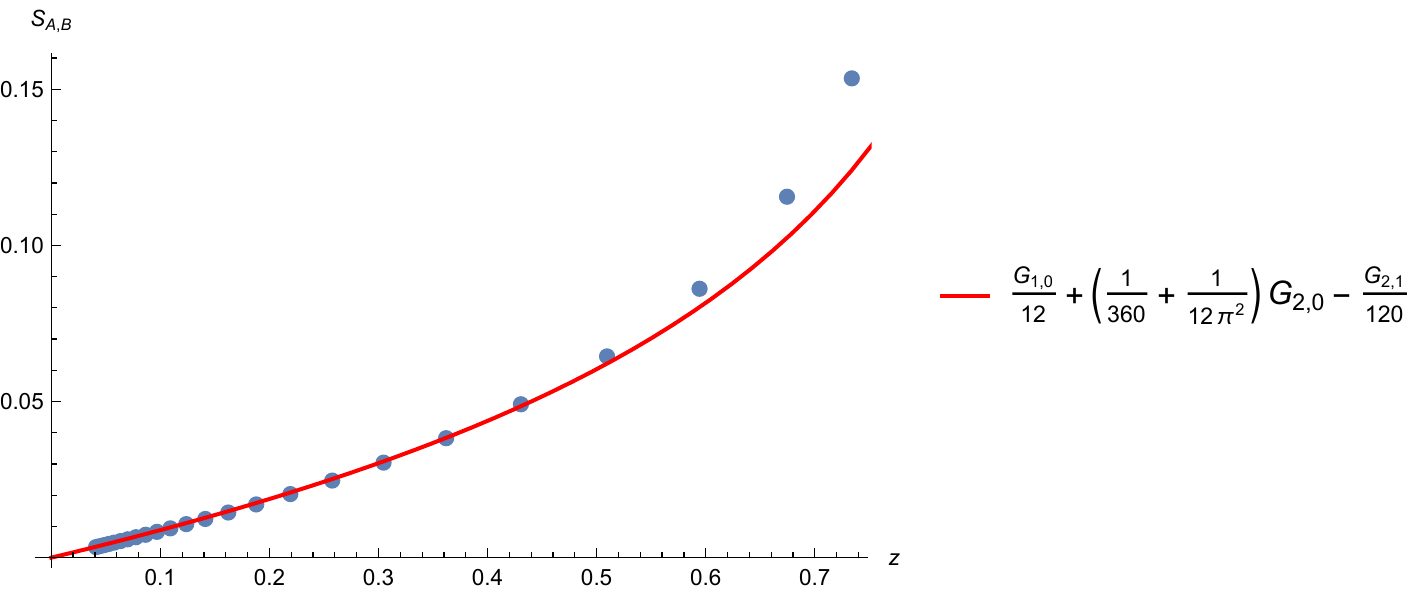}
     \caption{Mutual information of two disks for $3d$ free boson ($0<z<0.75$). }\label{mutual3dscalar1}
\end{figure}

\begin{figure}
\centering
 \includegraphics[width=0.5\textwidth]{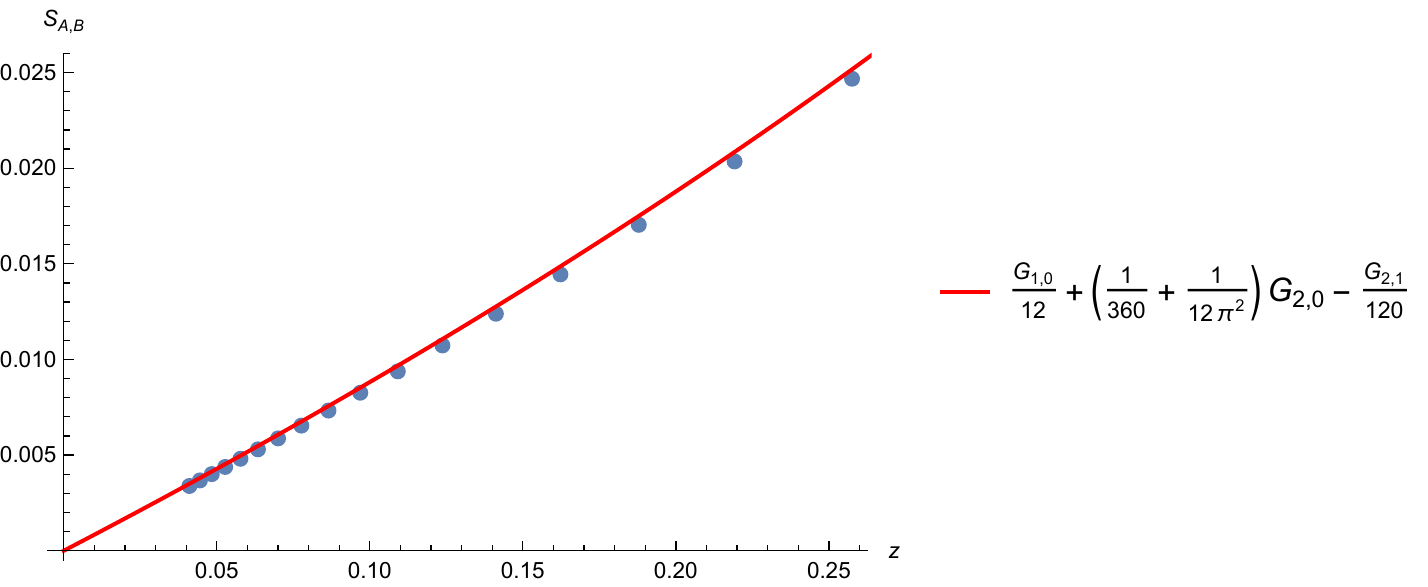}
     \caption{Mutual information of two disks for $3d$ free boson ($0<z<0.25$). }\label{mutual3dscalar2}
\end{figure}

 Figure \ref{mutual3dscalar1} shows the lattice computation and analytic result from the conformal block expansion for $3d$ free scalar.  
 Let us give some remarks on this result.
\begin{enumerate}
\item The conformal block expansion is very powerful as we have just computed the contribution from the first few operators with the low dimensions. The result is closer to the lattice computation in a wide region of $z$.  For example, $z$ is not restricted to $z\ll1$. Actually, in Figure \ref{mutual3dscalar1}  the parameter $0<z<0.75$ and the analytic result fits well in the region $0<z<0.55$. 
\item However, we expect the result (\ref{3dscalar}) to be modified significantly once $z\to 1$ as in this limit other conformal blocks may  be very important. 
\item The lattice result is not clear when $z<0.25$ in the figure. We zoom in this region in Figure \ref{mutual3dscalar2}.  As we expect, in this region,  the figure is well approximated by a line and the higher order correction can be neglected.  Then in this large distance limit, the mutual information is well approximated by $I_{A,B}=\frac{1}{12}z$.
\end{enumerate}

\begin{figure}
\centering
 \includegraphics[width=0.5\textwidth]{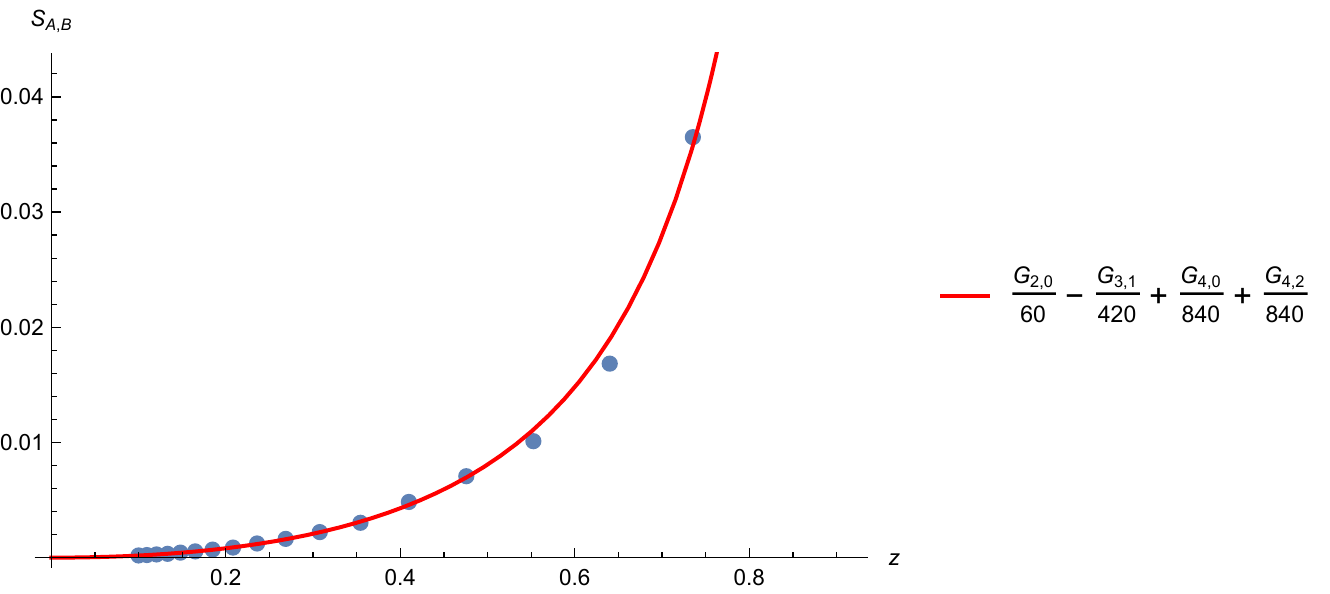}
     \caption{Mutual information of two balls for $4d$ free boson ($0<z<0.8$). }\label{mutual4dscalar1}
\end{figure}

\begin{figure}
\centering
 \includegraphics[width=0.5\textwidth]{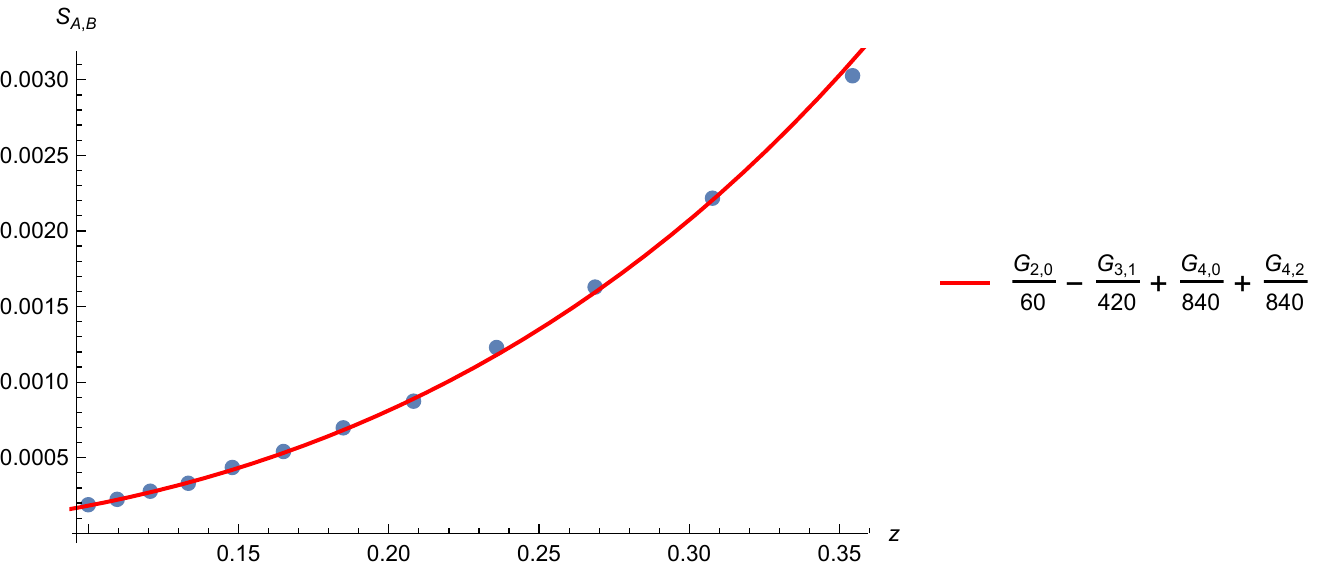}
     \caption{Mutual information of two balls for $4d$ free boson ($0<z<0.4$). }\label{mutual4dscalar2}
\end{figure}

In Figure \ref{mutual4dscalar1} and \ref{mutual4dscalar2}, we compare the lattice result and analytic result for $4d$ free scalar.  Note the analytic result (\ref{4dscalar}) fits very well in the region $0<z<0.8$.  Certainly, for even larger $z$, other conformal blocks  may contribute significantly.  

For higher dimensions with $d>4$, the operators of the first few lowest dimensions  are just of quadratic form.  We find the mutual information to be
\bea
I_{A,B}^{d>4}&=&\frac{4^{1 - d} \sqrt{\pi} \Gamma[d-1]}{\Gamma[-\frac{1}{2} + d]}G^{d>4}_{d-2,0}(z)-\frac{4^{-d} (d-2) \sqrt{\pi} \Gamma[d-1]}{\Gamma[\frac{1}{2} + d]}G^{d>4}_{d-1,1}(z)\nn\\&&+\frac{2^{-3 - 2 d} (4 -4d + 3 d^2)\sqrt{\pi} \Gamma[d-1]}{\Gamma[\frac{3}{2}+d]}G^{d>4}_{d,2}(z)+\cdots
\eea

\section{Free fermion theory}

For a generic fermionic operator, it has two-point correlation function
\be
<\psi_{\alpha}\psi^{\dag}_{\beta}>=\frac{\slashed{n}_{\alpha\beta}}{|x-x'|^{2\Delta}},
\ee
where
\be
\slashed{n}_{\alpha\beta}=\gamma^\mu_{\alpha\beta}(x-x')_{\mu}
\ee
with
\be
x_{\mu}=(r\sin\theta,r\cos\theta,y^i).
\ee
For a Dirac fermion, the independent number (real) of degrees of freedom is  $2^{[d/2]}\times 2$. The single fermionic operator does not have contribution to the mutual information. The leading order mutual information is contributed by the bilinear operators which are constructed from $\psi^{\dagger},\psi$ and $\gamma$ matrices. The total number of degrees of freedom of the bilinear operator is $2^{2[d/2]}$.   We can construct $d+1$ independent types of bilinear operators  in terms of $\psi,\bar{\psi}$ in even dimensions
\be
O^{(0)}=\psi^{\dagger}\psi,\ O^{(1)}_{\mu}=\psi^{\dag}\gamma_{\mu}\psi,\ O^{(2)}_{\mu\nu}=\psi^{\dag}[\gamma_{\mu},\gamma_{\nu}]\psi,\ \cdots, \ O^{(d)}=\psi^{\dagger}[\gamma_{\mu_1},[\gamma_{\mu_2},\cdots,\gamma_{\mu_d}]]\psi.
\ee
This can be shown by counting the total degrees of freedom
\be
\sum_{k=0}^dC_d^k=2^d
\ee
which is $2^{2[d/2]}$ for even dimensions. In odd dimensions, we can only construct $\frac{d+1}{2}$ independent types of bilinear operators
\be
O^{(0)}=\psi^{\dagger}\psi,\ O^{(1)}_{\mu}=\psi^{\dag}\gamma_{\mu}\psi,\ \cdots,\ O^{((d-1)/2)}_{\mu_1\mu_2\cdots\mu_{(d-1)/2}}=\psi^{\dag}[\gamma_{\mu_1},[\gamma_{\mu_2},\cdots,\gamma_{\mu_{(d-1)/2}}]]\psi.
\ee
Again one can count the degrees of freedom
\be
\sum_{k=0}^{\frac{d-1}{2}}C_d^k=2^{d-1}
\ee
which is exactly $2^{2[d/2]}$ in odd dimensions.
The first-type operator $O^{(0)}$ is a scalar, but it has no contribution to the mutual information.  The second-type operator  is of spin $1$, and it can be decomposed into two kinds: one of them is
\be
O_{\mu}^{(f)}=\psi^{\dag}_{j_1}\gamma_{\mu}\psi_{j_2}-(j_1\leftrightarrow j_2)\label{o}
\ee
with
\be
N^{(f)}_{2\Delta,1}=2 \tr 1_d,\ a^{(f)}_{2\Delta,1}=2(-1)^{j_1+j_2} \tr 1_d\frac{1}{(4s_{j_1j_2}^2)^{\Delta}}+\mathcal{O}(n-1),\label{a}
\ee
while the other one  is
\be
\tilde{O}_{\mu}^{(f)}=\psi^{\dag}_{j_1}\gamma_{\mu}\psi_{j_2}+(j_1\leftrightarrow j_2),\label{to}
\ee
whose one-point function is vanishing. There are two subtle points to derive (\ref{a}). Firstly, to use the $1/n$ prescription, the Green's function in the conifold geometry should be periodic.  For a fermionic theory,  its Green's function $G_{F}^{(n)}$ actually satisfies the following boundary condition \cite{Casini:2009sr}
\be
G_{F}^{(n)}(\theta+2\pi n)=(-1)^{n-1}G_F^{(n)}(\theta).
\ee
Only when $n$ is odd, the Green's function is periodic. We expect odd $n$ result is already enough to determine the coefficient $b_{2\D,1}$. The other subtle point is
that we have included a factor of $(-1)^{j_1+j_2}$ for the coefficient $a_{2\Delta,1}$ as there will be a factor $(-1)$ for the fermion once it rotates $2\pi$.  This can be shown explicitly for free fermions below.
In $d$ dimension, $\tr 1_d=2^{[\frac{d}{2}]}$, then the contribution from the spin $1$ operator is
\be
b^{(f)}_{2\Delta,1}=\frac{2^{[\frac{d}{2}]-1}\sqrt{\pi}\Gamma[2\Delta+1]}{4^{2\Delta}\Gamma[2\Delta+\frac{3}{2}]}.\label{fermionspin1}
\ee
 Note that the coefficient is positive which is different in sign from the one constructed from scalar operator. 
Physically this coefficient must be positive as it gives the leading order mutual information.
The other operators are antisymmetric in their indices so that they have no contribution to the mutual information.

For a free Dirac fermion in $d$ dimension,  we have $\Delta=\frac{d-1}{2}$.  We list the leading order  mutual information in various dimensions
\bea
I_{d=2}&=&\frac{1}{3}G^{(d=2)}_{1,1}(z)+\cdots=\frac{1}{3}z+\cdots,\label{d2}\\
I_{d=3}&=&\frac{1}{15}G^{(d=3)}_{2,1}(z)+\cdots=\frac{1}{15}z^2+\cdots,\label{d3}\\
I_{d=4}&=&\frac{1}{35}G^{(d=4)}_{3,1}(z)+\cdots=\frac{1}{35}z^3+\cdots.\label{d4}
\eea
Let us check the above result case by case. In $2d$, the exact mutual information is of extensive nature and given by \cite{Casini:2009sr}
\be
I_{d=2}=-\frac{1}{3}\log(1-z)=\frac{1}{3}z+\cdots
\ee
which is in good match with $(\ref{d2})$. Actually, the conformal block is exactly
\be
G^{(d=2)}_{1,1}(z)=-\log(1-z).
\ee
This indicates that the other primary modules in the replicated theory should cancel each other, even though each of them may have non-vanishing contribution. 

\begin{figure}
\centering
 \includegraphics[width=0.5\textwidth]{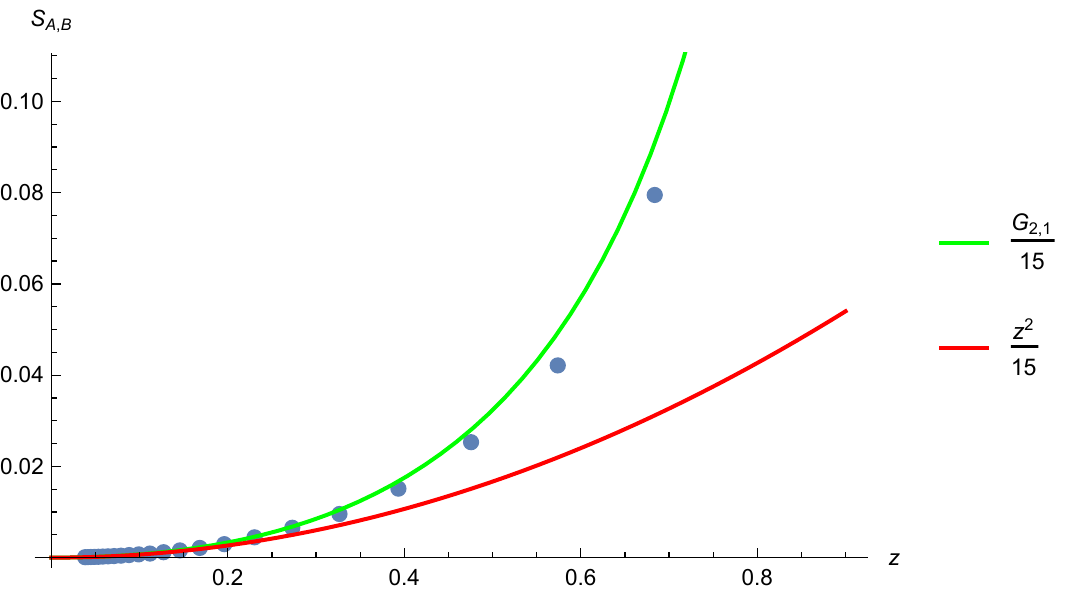}
     \caption{Mutual information of two disks for $3d$ free Dirac fermion, $0<z<0.9$. The red line is the leading order contribution in $z$, while the green line is the leading order conformal block.}\label{mutual3dfermion}
\end{figure}

\begin{figure}
\centering
 \includegraphics[width=0.5\textwidth]{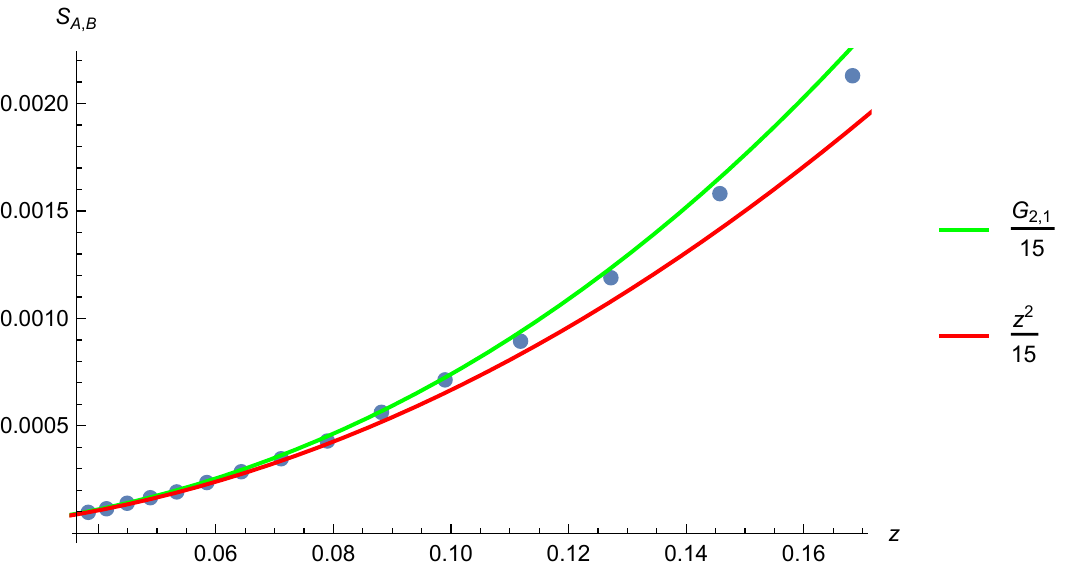}
     \caption{Mutual information of two disks for $3d$ free Dirac fermion, $0<z<0.17$. The red line is the leading order contribution in $z$, while the green line is the leading order conformal block.}\label{mutual3dfermion1}
\end{figure}

\begin{figure}
\centering
 \includegraphics[width=0.5\textwidth]{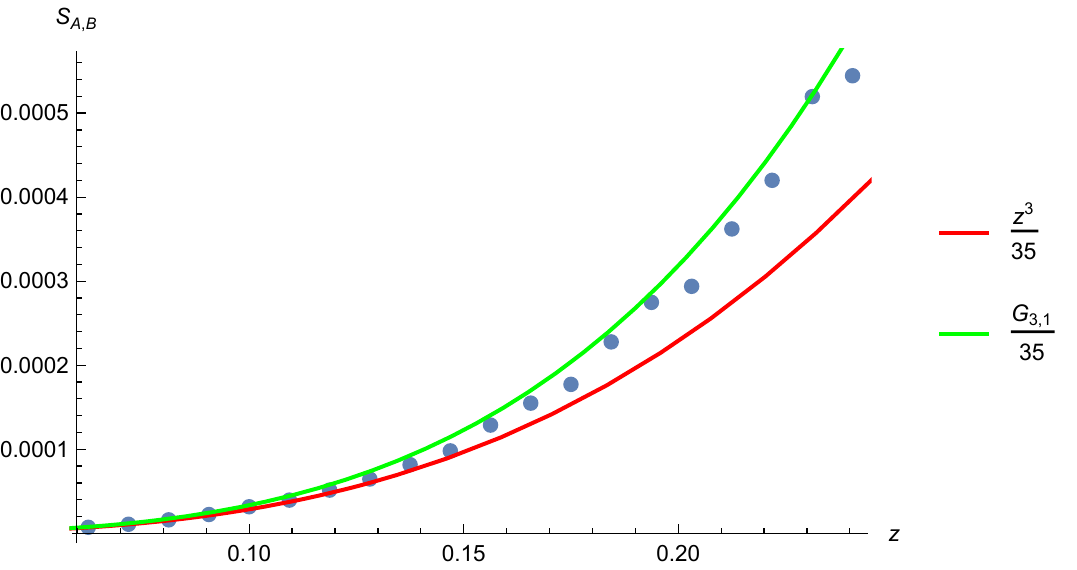}
     \caption{Mutual information of two balls for $4d$ free Dirac fermion, $0<z<0.25$. The red line is the leading order contribution in $z$, while the green line is the leading order conformal block. }\label{mutual4dfermion}
\end{figure}

For the free Dirac fermions in higher dimensions, we can compare the above analytic result with the numerical computation, which has been discussed in Appendix \ref{numerical}.  We compare the numerical result and the leading analytic result for $3d$ free Dirac fermion in Figure \ref{mutual3dfermion}. It is easy to see that the conformal block expansion has better match when $z$ is not very small. Figure \ref{mutual3dfermion1} is the same date while  restricting to small $z$ region.
In $4d$, we can also compute the mutual information of free fermion numerically. In Figure \ref{mutual4dfermion}, we compare the numerical result and the leading analytic result.
Once again, we find the better match from the conformal block expansion when $z$ is not very small.

The above analysis on the fermionic type operator comes from the $1/n$ prescription. It would be better to have a direct computation. 
We can study the free Dirac fermion in a general dimension. The two-point function in the conifold geometry can be obtained by using the image method for odd $n$, after taking into account of the spin structure carefully\cite{Herzog:2015cxa}
\bea
<\psi(x)\psi^\dagger(x')>_{\frac{1}{n}}&=&-\frac{1}{d-2}\gamma^{\mu}\partial_{\mu}\sum_{k=0}^{n-1}\frac{(-1)^k(\cos\frac{\pi k}{n}+\gamma^0\gamma^1\sin\frac{\pi k}{n})}{(r^2+r'^2-2r r'\cos(\theta-\theta'-\frac{2\pi k}{n}))^{\frac{d-2}{2}}}\nn\\&=&-\frac{1}{d-2}\gamma^{\mu}\partial_{\mu}\frac{1}{(4rr')^{\frac{d-2}{2}}}\sum_{k=0}^{n-1}\frac{(-1)^k(\cos\frac{\pi k}{n}+\gamma^0\gamma^1\sin\frac{\pi k}{n})}{(\sinh(\eta^+-i\frac{\pi k}{n})\sinh(\eta^-+i\frac{\pi k}{n}))^{\frac{d-2}{2}}}
\eea
where we have defined
\be
\cosh\eta=\frac{r^2+r'^2+(y-y')^2}{2rr'},\hs{2ex} \eta^{\pm}=\frac{\eta\pm i(\theta-\theta')}{2},
\ee
and assumed $\theta>\theta'$ in the discussion below.
As the analytic continuation commutes with the derivative, the one-point function coefficient of the spin-one operator $\psi^{\dagger}_{j_2}\gamma^{\mu}\psi_{j_1}$ can be obtained by computing
\bea
<\psi^\dagger(r,\theta',0)\gamma^0\psi(r,\theta,0)>_{\frac{1}{n}}&=&-\tr 1_d\sum_{k=0}^{n-1}\frac{(-1)^k\cos\frac{\theta+\theta'}{2}}{(2r\sin(\frac{\theta-\theta'}{2}-\frac{\pi k}{n}))^{d-1}}\nn\\&=&-\tr 1_d\frac{\cos\frac{\theta+\theta'}{2}}{(2r)^{d-1}}\sum_{k=0}^{n-1}(-1)^k\int_0^{\infty}dx x^{\frac{\theta-\theta'}{2\pi}-\frac{k}{n}-1}f_{d-1}(x)\nn\\&=&-\tr 1_d\frac{\cos\frac{\theta+\theta'}{2}}{(2r)^{d-1}}\int_0^{\infty}dx x^{\frac{\theta-\theta'}{2\pi}-1}f_{d-1}(x)\frac{1-(-1)^n x^{-1}}{1+x^{-\frac{1}{n}}}.\nn
\eea
As $n$ is odd, then the above identity can analytically continue from $\frac{1}{n}$ to $n$,
\bea
<\psi^\dagger(r,\theta',0)\gamma^0\psi(r,\theta,0)>_n&=&-\tr 1_d\frac{\cos\frac{\theta+\theta'}{2}}{(2r)^{d-1}}\int_0^{\infty}dx x^{\frac{\theta-\theta'}{2\pi}-1}f_{d-1}(x)\frac{1+ x^{-1}}{1+x^{-n}}\nn\\&=&-\tr 1_d\frac{\cos\frac{\theta+\theta'}{2}}{(2r)^{d-1}}\int_0^{\infty}dxx^{\frac{\theta-\theta'}{2\pi n}-1}f_{d-1}(x)+\mathcal{O}(n-1).\nn
\eea
Then we can read the one-point function coefficient
\be
a_{d-1,1}^{j_2j_1}=\tr 1_d\frac{1}{2^{d-1}s_{j_1j_2}^{d-1}}(-1)^{j_1+j_2+1}+\mathcal{O}(n-1),\ j_1>j_2.
\ee
Note by exchanging $j_1$ and $j_2$, there is an extra minus sign so
the one-point function coefficient of operator $O^{(f)}_{\mu}$ is the same as (\ref{a}).
The coefficient  in the mutual information
\be
b^{(f)}_{d-1,1}=2^{[\frac{d}{2}]+1-2d}\sqrt{\pi}\frac{\Gamma(d)}{\Gamma(d+\frac{1}{2})}
\ee
which is exactly the same as (\ref{fermionspin1}) after taking $d=2\D+1$. The leading order mutual information for a free fermion in $d$ dimension is just
\be
I_{d}=b^{(f)}_{d-1,1}G^{(d)}_{d-1,1}(z)+\cdots.
\ee

\section{Conclusions and discussions}

In this paper, we studied the mutual information of two disjoint spheres in a general conformal field theory. By considering the operator product expansion of the spherical twist operators in terms of the conformal family, we expanded the mutual information in terms of the conformal blocks. One essential point is that we should consider the primary operators in the replicated theory carefully. For simplicity, we only focused on the contribution from the conformal family $O$ of the lowest scaling dimension in the original theory. In this case, it turns out the single operator $O$ does not contribute to the mutual information, and the leading contribution comes from the bilinear operator $O_{j_1}O_{j_2}$ or its cousin in the replicated theory. The difficult part is to determine the coefficient of such bilinear operator in the OPE and the coefficient before the conformal block in the final expression. This requires us to compute the one-point function of the operator in the conifold geometry. Even though the exact form of the one-point function depends on the details of the theory and is therefore generically unknown, one can do expansion with respect to $(n-1)$ just in order to read the mutual information. This leads us to propose the so-called $1/n$ prescription, which suggests that the coefficients we are interested in depend actually only on the form of the two-point function of $O$ in the flat spacetime. Or in other words, the coefficient is determined by the behavior of the  operator $O$ under the conformal symmetry transformation. Thus the contribution of the bilinear operator $O_{j_1}O_{j_2}$ to the mutual information takes a universal form, in the sense that it is determined by the conformal symmetry. The similar discussion can be applied to the operators of bilinear forms involving the partial derivatives and/or the gamma matrices in the case of the fermionic operators, by using the $1/n$ prescription developed.

As an application of the prescription, we took the free scalar and free fermion as examples to illustrate the power of our approach. We computed the mutual information of the free scalars in three and four dimensions to the next-to-leading order and found quite good agreement with the numerical lattice computation. For the free fermion, the analytic computation of mutual information has not been done before our work. We analytically computed the leading order contribution in free fermion theories in various dimensions and found good agreement with the numerical results in 3D and 4D. Especially we found that the contributions in terms of the conformal blocks had better fitting than the simple leading order contributions in $z$, showing the advantage of the approach.

 The formalism presented in this work is quite general and powerful. In this work, we only focused on the single conformal family with the lowest dimension, it is straightforward to consider the case with many conformal families. The complication may come from two parts: one is that the construction of the primary operators in the replicated theory could be more complex, and the other is that we should take into account of the OPE of the conformal families. For example, suppose that there are two conformal families $O^{(i)}, i=1,2$ in a CFT. In the replicated theory, besides the primary operators constructed purely from $O^{(1)}$ or $O^{(2)}$, there could be primary operators composed of $O^{(1)}$ and $O^{(2)}$. The bilinear one like $O^{(1)}_{j_1}O^{(2)}_{j_2}$ has no contribution to the mutual information, but the quartic ones $ O^{(1)}_{j_1}O^{(2)}_{j_2}O^{(1)}_{j_3}O^{(2)}_{j_4}$ may have nonvanishing contribution. Such contribution is of higher order. In the case that there are three conformal families with OPE: $[O^{(1)}]\times [O^{(2)}] =C^3_{12}[O^{(3)}]$, the primary operator $O^{(1)}_{j_1}O^{(1)}_{j_2}O^{(1)}_{j_3}$ may have nonvanishing contribution as well.   Nevertheless, once the data of a CFT including the spectrum and the three-point coefficients is known, the mutual information could be computed in principle in the framework.

For the free fermion theory, even it is simple, the analytic study of the mutual information in this work is new. One interesting question is to consider the next-to-leading contributions to the mutual information. It is also interesting to discuss the R\'enyi mutual information, which could be more challenging.

Another interesting application of our work is to the holographic theory. In this case, we should include the conformal family of the stress tensor. In the semiclassical AdS$_3$/CFT$_2$ correspondence, the  CFT is of large central charge and the correlation function of the twist operators is dominated by the vacuum module, which is mainly made of the stress tensor and its derivatives. At the leading order, the contribution from the bilinear operators of the stress tensor is given by (\ref{spin2}).  In $d=2$ only the spin 4 part contributes $b_{4,4}=\frac{1}{630}$ which is the same as the one got in \cite{Chen:2013kpa}. In the other cases of AdS/CFT,  there are other conformal families besides the stress tensor in CFT such that the holographic study of the mutual information is more challenging. In our study, we showed that the bilinear operators including the partial derivatives contribute to the mutual information universally, how to show this fact and read the universal behavior holographically is an interesting question.

The OPE expansion of the twist operator can be applied not only to the mutual information of disjoint spheres, but also to other problems. One possible application is to the thermal and finite size correction to the entanglement entropy. In 2D large $c$ CFT, this technique has been applied to the single interval on a torus\cite{Chen:2014unl}. It would be interesting to pursue this issue in higher dimensions\cite{Herzog:2014fra,Herzog:2014tfa}.

\section*{Acknowledgments}

B.C. was in part supported by NSFC Grant No.~11275010, No.~11335012 and No.~11325522. BC would like to thank the hospitality of Osaka City University, where the final stage of this work was finished. J.L. is supported by the ERC Starting Grant 335146 ``HoloBHC".

\appendix
\section{Conventions}

In this appendix, we list the convention we used in this paper.
\subsection{Projector}

We used a few projectors in our study. They are defined as follows. In the spacetime of dimension $d$, the spin 2 projector is of the form
\be
P^{\nu_1\nu_2}_{\mu_1\mu_2}=\frac{1}{2}\delta^{\nu_1}_{(\mu_1}\delta^{\nu_2}_{\mu_2)}-\frac{1}{d}\delta_{\mu_1\mu_2}\delta^{\nu_1\nu_2},
\ee
and the spin 3 projector is of the form
\be
P^{\nu_1\nu_2\nu_3}_{\mu_1\mu_2\mu_3}=\frac{1}{6}\delta^{\nu_1}_{(\mu_1}\delta^{\nu_2}_{\mu_2}\delta^{\nu_3}_{\mu_3)}-\frac{1}{3(d+2)}\delta_{(\mu_1\mu_2}\delta^{(\nu_1\nu_2}\delta^{\nu_3)}_{\mu_3)},
\ee
and the spin 4 projector is
\be
P^{\nu_1\nu_2\nu_3\nu_4}_{\mu_1\mu_2\mu_3\mu_4}=\frac{1}{24}\delta^{\nu_1}_{(\mu_1}\delta^{\nu_2}_{\mu_2}\delta^{\nu_3}_{\mu_3}\delta^{\nu_4}_{\mu_4)}-\frac{1}{12(d+4)}\delta_{(\mu_1\mu_2}\delta^{(\nu_1\nu_2}\delta_{\mu_3}^{\nu_3}\delta_{\mu_4)}^{\nu_4)}+\frac{1}{3(d+2)(d+4)}\delta_{(\mu_1\mu_2}\delta_{\mu_3\mu_4)}\delta^{(\nu_1\nu_2}\delta^{\nu_3\nu_4)}.
\ee

\subsection{One point functions on conifold geometry}
By symmetry consideration\cite{Long:2016vkg}, the one-point functions of various operators in the conifold geometry are determined up to coefficients. Here are the forms of the nonvanishing one-point functions of the operators from spin 0 to spin 4
\bea
&&<O(x)>_n=a_{\Delta,0}\frac{1}{|x|^{\Delta}},\hs{3ex}
<O_a>_n=a_{\Delta,1}\frac{\epsilon_{ab}n^b}{|x|^{\Delta}},\nn\\
&&<O_{ab}>_n=-a_{\Delta,2}\frac{(d-1)\delta_{ab}-dn_an_b}{|x|^{\Delta}},\hs{3ex} <O_{ij}>_n=a_{\Delta,2}\frac{\delta_{ij}}{|x|^{\Delta}}\nn\\
&&<O^{abc}>_n=a_{\Delta,3}\frac{1}{3}\frac{((1-d)\delta^{ab}+(d+2)n^an^b)\epsilon^{cd}n_d+perm}{|x|^{\Delta}}\nn\\
&&<O^{aij}>_n=a_{\Delta,3}\delta^{ij}\frac{\epsilon^{ab}n_b}{|x|^{\Delta}}\nn\\
&&<O_{abcd}>_n=a_{\Delta,4}\frac{(d+2)(d+4)n_an_bn_cn_d-\frac{(d+1)(d+2)}{3}\delta_{(ab}n_cn_{d)}+\frac{d^2-1}{3}\delta_{(ab}\delta_{cd)}}{|x|^{\Delta}},\nn\\
&&<O_{abij}>_n=a_{\Delta,4}\delta_{ij}\frac{-(d+1)\delta_{ab}+(d+2)n_an_b}{|x|^{\Delta}},\hs{3ex} <O_{ijkl}>_n=a_{\Delta,4}\frac{\delta_{(ij}\delta_{kl)}}{|x|^{\Delta}}.\nn
\eea

\subsection{Two point functions on flat spacetime}
The two-point function of a primary operator of an integer spin $s$ in the flat spacetime is of the form
\be
<O(x)_{\mu_1\cdots\mu_s}O(x')_{\nu_1\cdots\nu_s}>=P^{\alpha_1\cdots\alpha_s,\beta_1\cdots\beta_s}_{\mu_1\cdots\mu_s,\nu_1\cdots\nu_s}\frac{I_{\alpha_1\beta_1}\cdots I_{\alpha_s\beta_s}}{|x-x'|^{2\Delta}},
\ee
where the projector $ P^{\alpha_1\cdots\alpha_s,\beta_1\cdots\beta_s}_{\mu_1\cdots\mu_s,\nu_1\cdots\nu_s}$ is to project the quantity $I_{\alpha_1\beta_1}\cdots I_{\alpha_s\beta_s}$ to be symmetric and traceless in the indices $\alpha$ and $\beta$ separately. We list the two-point functions of the operators from spin 0 to spin 4 explicitly below
\bea
<O(x)O(x')>&=&N_{\Delta,0}\frac{1}{|x-x'|^{2\Delta}},\nn\\
<O_{\mu}(x)O_{\nu}(x')>&=&N_{\Delta,1}\frac{I_{\mu\nu}(x-x')}{|x-x'|^{2\Delta}},\nn\\
<O_{\mu_1\mu_2}O_{\nu_1\nu_2}(x')>&=&N_{\Delta,2}\frac{\frac{1}{2}(I_{\mu_1\nu_1}I_{\mu_2\nu_2}+I_{\mu_1\nu_2}I_{\mu_2\nu_1})-\frac{1}{d}\delta_{\mu_1\mu_2}\delta_{\nu_1\nu_2}}{|x-x'|^{2\Delta}},\nn\\
<O_{\mu_1\mu_2\mu_3}(x)O_{\nu_1\nu_2\nu_3}(x')>&=&N_{\Delta,3}\frac{Q_{\mu_1\mu_2\mu_3,\nu_1\nu_2\nu_3}^{(3)}}{|x-x'|^{2\Delta}},\nn\\
<O_{\mu_1\mu_2\mu_3\mu_4}(x)O_{\nu_1\nu_2\nu_3\nu_4}(x')>&=&N_{\Delta,4}\frac{Q_{\mu_1\mu_2\mu_3\mu_4,\nu_1\nu_2\nu_3\nu_4}^{(4)}}{|x-x'|^{2\Delta}},\nn
\eea
where
\bea
Q_{\mu_1\mu_2\mu_3,\nu_1\nu_2\nu_3}^{(3)}&=&\frac{1}{6}(I_{\mu_1\nu_1}I_{\mu_2\nu_2}I_{\mu_3\nu_3}+\mbox{perm})
-\frac{1}{3(d+2)}(\delta_{\mu_1\mu_2}\delta_{\nu_1\nu_2}I_{\mu_3\nu_3}+\mbox{perm}),\nn\\
Q_{\mu_1\mu_2\mu_3\mu_4,\nu_1\nu_2\nu_3\nu_4}^{(4)}&=&\frac{1}{24}(I_{\mu_1\nu_1}I_{\mu_2\nu_2}I_{\mu_3\nu_3}I_{\mu_4\nu_4}+\mbox{perm})
-\frac{1}{12(d+4)}(\delta_{\mu_1\mu_2}\delta_{\nu_1\nu_2}I_{\mu_3\nu_3}I_{\mu_4\nu_4}+\mbox{perm})\nn\\
&&+\frac{1}{3(d+2)(d+4)}(\delta_{\mu_1\mu_2}\delta_{\mu_3\mu_4}\delta_{\nu_1\nu_2}\delta_{\nu_3\nu_4}+\mbox{perm}).\nn
\eea

\subsection{Summation}

In order to do various summations involving the inverse of the function $\sin^{\Delta}{\frac{\theta}{2}}$, we use the following integral representation
\be
\frac{1}{\sin^{\Delta}{\frac{\theta}{2}}}=\int_0^{\infty}dx x^{\frac{\theta}{2\pi}-1}f_{\Delta}(x),
\ee
where
\be
f_{\Delta}(x)=\frac{2^{\Delta-2}}{\pi^2\Gamma[\Delta]\sqrt{x}}\Gamma[\frac{\Delta}{2}+i\frac{\log x}{2\pi}]\Gamma[\frac{\Delta}{2}-i\frac{\log x}{2\pi}].
\ee
Then we can obtain the following summations
\bea
\sum_{j=1}^{n-1}\frac{1}{s_{j}^{\Delta}}&=&(n-1)\frac{\sqrt{\pi}}{2}\frac{\Gamma[\frac{\Delta}{2}+1]}{\Gamma[\frac{\Delta+3}{2}]}+\mathcal{O}(n-1)^2\\
\sum_{j_1\not=j_2>j3}\frac{1}{s_{j_1j_2}^{\Delta}s_{j_1j_3}^{\Delta}}&=&-\frac{\sqrt{\pi}}{4}\frac{\Gamma[\Delta+1]}{\Gamma[\Delta+\frac{3}{2}]}(n-1)+\mathcal{O}(n-1)^2
\eea


\section{The coefficients $f_{\D,J}$}
The coefficients $f_{\D,J}$ can be obtained following \cite{Long:2016vkg}. Here we present another way to fix their values. Let us focus on the primary operator $\mathcal{O}_{\Delta,J}$ in the replicated theory. We work in the Euclideanized theory, and find
\bea
\tr_A\;\rho_A^n=\frac{Z_n}{Z_1^n}.
\eea
 Here $Z_1$ is the Euclidean partition function on $R^d$ and $Z_n$ is the Euclidean partition function on $\mathcal{C}^n_A$ which is a manifold of $n$ copied $R^{d}$ glued together along the ball region $A$.

As we stated before, the contribution from the primary module ${O}_{\Delta,J}$  to the partition function could be written as
\bea
R_{\Delta,J}
&\equiv &c_{\Delta,J}(R_A)c_{\Delta,J}(R_B)\langle Q[{O}_{\Delta,J}(x_A)]Q[{O}_{\Delta,J}](x_B)\rangle\nn\\
&=&c_{\Delta,J}(R_A)c_{\Delta,J}(R_B)\langle {O}_{\Delta,J}(x_A){O}_{\Delta,J}(x_B)\rangle+ \mbox{subleading terms},
\eea
when $|x_A-x_B|$ is large. The first term on right hand side is actually a shorthand of
\bea
R^0_{\D,J}=c_{\Delta,J}(R_A)c_{\Delta,J}(R_B)\langle a^{\mu_1\mu_2\dots\mu_J}{O}_{\mu_1\mu_2\dots\mu_J}(x_A)a^{\nu_1\nu_2\dots\nu_J}{O}_{\nu_1\nu_2\dots\nu_J}(x_B)\rangle,
\eea
where $a^{\mu_1\mu_2\dots\mu_J}$ is a tensor independent of position and is to be determined.  To be more concrete, we choose our convention as
\begin{eqnarray}
  \langle{O}_{\mu_1\mu_2\dots\mu_J}(x){O}_{\nu_1\nu_2\dots\nu_J}(x^\prime)\rangle =N_{\Delta,J}\frac{Q_{\mu_1\mu_2\dots\mu_J,\nu_1\nu_2\dots\nu_J}(n^\mu_{xx^\prime})}{|x-x^\prime|^{2\Delta}},
\end{eqnarray}
where
\be
n^\mu_{xx^\prime}=(x^\mu-x^{\prime\mu})/|x-x^\prime|
 \ee
 and $Q_{\mu_1\mu_2\dots\mu_J,\nu_1\nu_2\dots\nu_J}(n^\mu_{xx^\prime})$ is conformal invariant tensor constructed by $n^\mu_{xx^\prime}$ and $g_{\mu\nu}$ ($\delta_{\mu\nu}$ if choosing the Cartesian coordinates). Now we can write $R^0_{\D,J}$ as
\bea
R^0_{\D,J}=c_{\Delta,J}(R_A)c_{\Delta,J}(R_B)a^{\mu_1\mu_2\dots\mu_J}a^{\nu_1\nu_2\dots\nu_J}
N_{\Delta,J}\frac{Q_{\mu_1\mu_2\dots\mu_J,\nu_1\nu_2\dots\nu_J}(n^\mu_{x_Ax_B})}{|x_A-x_B|^{2\Delta}}.
\eea
Under the rotation, $Q_{\mu_1\mu_2\dots\mu_J,\nu_1\nu_2\dots\nu_J}$ is a (0,$2J$) tensor, while $R^0_{\D,J}$ is a scalar so that $a^{\mu_1\mu_2\dots\mu_J}$ must be a ($J$,0) tensor. Suppose we have chosen a specific Cartesian coordinate, then $a^{\mu_1\mu_2\dots\mu_J}$ is fixed and we can read out $n^\mu_{x_Ax_B}$ too. For simplicity we always have $x^0_A=x^0_B$, so $n^0_{x_Ax_B}=0$.

 The scalar $R_{\Delta,J}$ is closely related to mutual information, and it is dependent of  $n^i_{x_Ax_B}$. When we rotate $n^i_{x_Ax_B}$, $R^0_{\D,J}$ does not change. We can choose $n^\mu_{x_Ax_B}=n^\mu_s\equiv(0,1,0,\dots,0)$ without lose of generality.
Under this $SO(d-1)$ rotation, the tensor  $Q_{\mu_1\mu_2\dots\mu_J,\nu_1\nu_2\dots\nu_J}(n^\mu_{x_Ax_B})$
behaves  as the tensor under the rotation  $x^i\to R^i_{\;\;j}x^j$, where $R^i_{\;\;j}$ is a $SO(d-1)$ rotation matrix. As $R^0_{\D,J}$ is invariant, the quantity $a^{\mu_1\mu_2\dots\mu_J}$ should obey the following constraint
\begin{eqnarray}
  a^{\mu_1\mu_2\dots\mu_J} &=& R^{\mu_1}_{\;\;\;\nu_1}R^{\mu_2}_{\;\;\;\nu_2}\dots R^{\mu_J}_{\;\;\;\nu_J}a^{\nu_1\nu_2\dots\nu_J}, \\
  \mbox{where}\;\;R^0_{\;\;0} &=& 1,\;\;\; R^i_{\;\;0}=R^0_{\;\;i}=0.\nn
\end{eqnarray}
This can be used to determine the form of $a^{\mu_1\mu_2\dots\mu_J}$. For example, for a vector type primary operator, we have
\bea
a^{\mu}=R^{\mu}_{\;\;\;\nu}a^{\nu}, \;\;\;\forall R^i_{\;\;j}.
\eea
Then we get $a^i=0$. Similarly, for a rank-2 symmetric tensor we can get
\bea
a^{00}=a_1,\;\;\;a^{0i}=0,\;\;\;a^{ij}=a_2\;\delta^{ij}.
\eea

 The one-point function of ${O}_{\Delta,J}$ on $\mathcal{C}^n_A$ could be understood as the correlation function between the operator and the defect operator. If the primary operators have been normalized and orthogonalized properly,  the one-point function is actually
\begin{equation}
  \langle{O}_{\Delta,J}(x)\rangle_{\mathcal{C}^n_A}= \langle{O}_{\Delta,J}(x)c_{\Delta,J}(R_A)Q[{O}_{\Delta,J}](x_A)\rangle,
\end{equation}
where $x_A$ is the position of the center of $A$. More explicitly
\begin{eqnarray}
  \langle{O}_{\mu_1\mu_2\dots\mu_J}(x)\rangle_{\mathcal{C}^n_A} &=& \langle{O}_{\mu_1\mu_2\dots\mu_J}(x)c_{\Delta,J}(R_A)a^{\nu_1\nu_2\dots\nu_J}{O}_{\nu_1\nu_2\dots\nu_J}(x_A)\rangle+\dots \nn\\
   &=& c_{\Delta,J}(R_A)a^{\nu_1\nu_2\dots\nu_J}\langle{O}_{\mu_1\mu_2\dots\mu_J}(x){O}_{\nu_1\nu_2\dots\nu_J}(x_A)\rangle+\dots\nn\\
&=&c_{\Delta,J}(R_A)a^{\nu_1\nu_2\dots\nu_J}N_{\Delta,J}\frac{Q_{\mu_1\mu_2\dots\mu_J,\nu_1\nu_2\dots\nu_J}(n^\mu_{xx_A})}{|x-x_A|^{2\Delta}}+\dots.
\end{eqnarray}
Now we further set $\mu_1=\mu_2=\dots=\mu_J=0$ for simplicity.  And we can choose a specific $n^\mu_{xx_A}=n^\mu_s\equiv(0,1,0,\dots,0)$ without lose of generality. So we arrive at
\bea
\langle{O}_{00\dots0}(0,x^i)\rangle_{\mathcal{C}^n_A} =
c_{\Delta,J}(R_A)a^{\nu_1\nu_2\dots\nu_J}N_{\Delta,J}\frac{Q_{00\dots0,\nu_1\nu_2\dots\nu_J}(n^\mu_s)}{|x^i-x^i_A|^{2\Delta}}+\dots.
\eea
When $|x^i-x^i_A|$ is large, the first term on the right hand side is the leading term. So we can determine $c_{\Delta,J}(R_A)$ as following
\begin{eqnarray}
  c_{\Delta,J}(R_A) &=& \lim_{x^i\to \infty}\frac{\langle{O}_{00\dots0}(0,x^i)\rangle_{\mathcal{C}^n_A} |x^i-x^i_A|^{2\Delta}}{N_{\Delta,J}a^{\nu_1\nu_2\dots\nu_J}Q_{00\dots0,\nu_1\nu_2\dots\nu_J}(n^\mu_s)}.
\end{eqnarray}

In the above discussion,  $\mathcal{C}^n_A$ is a manifold of $n$-copied $R^{d}$ with the ball region $A$ being glued. It turns out to be more convenient to study the one-point function in  the planar conifold geometry. The planar conifold geometry is related to the spherical conical geometry by an inversion map. We define $\widetilde{\mathcal{C}}^n_A$ to be a manifold of $n$-replicated  $R^{d}$ with a half plane being glued.  For the ball $A$, we may put its origin at $(t_E,x^1=R_A,0,0,\ldots)$ in the Cartesian coordinates. Under the inversion $x^\mu\to x^\mu/x^2$, the surface of $A$ at $t_E=0$  becomes a plane located at $x^1=1/(2R_A),\;t_E=0$. And the replicated plane is of a conical singularity. The points at the infinity is mapped to (0,0,0,\ldots). Its distance to the singularity is $1/(2R_A)$.

 Under a conformal transformation $x\to x'$, the operators $\cO_{\D,J}(x)$ can be related to the  operators $\widetilde{\cO}_{\D,J}(x')$
\begin{equation}
\widetilde{{O}}_{\Delta,J}(x^\prime)=b(x)^{-\Delta}\cR[I^\mu_{\;\;\nu}(x)]{O}_{\Delta,J}(x),
\end{equation}
where $b(x)$ and $I^\mu_{\;\;\nu}(x)$ are defined by
\begin{equation}
  \frac{\partial x^{\prime\mu}}{\partial x^\nu}=b(x)I^\mu_{\;\;\nu}(x),\;\;\;\det(I^\mu_{\;\;\nu}(x))=\pm1,\;\;b(x)>0,
\end{equation}
and $\cR[I^\mu_{\;\;\nu}(x)]$ is some representation of $O(d)$. For the inversion $x^\mu\to x^{\prime\mu}=x^\mu/x^2$, we have
\begin{equation}
  b(x)=x^{-2},\;\;\;\;I^\mu_{\;\;\nu}(x)=\delta^\mu_{\;\;\nu}-2n^\mu n_\nu,\;\;\;\;n^\mu=\frac{x^\mu}{|x|}.
\end{equation}
And then we get
\begin{equation}
\widetilde{{O}}_{\Delta,J}(x^\mu/x^2)=x^{2\Delta}\cR[I^\mu_{\;\;\nu}(x^\mu)]{O}_{\Delta,J}(x^\mu)
\end{equation}
which is the shorthand writing of
\bea
\widetilde{{O}}_{\mu_1\mu_2\dots\mu_J}(x^\mu/x^2)=x^{2\Delta}I^{\nu_1}_{\;\;\mu_1}(x^\mu)I^{\nu_2}_{\;\;\mu_2}(x^\mu)\dots I^{\nu_J}_{\;\;\mu_J}(x^\mu){O}_{\nu_1\nu_2\dots\nu_J}(x^\mu).
\eea
When $x^\mu=(0,x^i)$, we have $I^{\nu}_{\;\;0}(0,x^i)= \delta^{\nu}_{\;\;0}$, so
\bea
\widetilde{{O}}_{00\dots0}(0,x^i/x^2)=x^{2\Delta} {O}_{00\dots0}(0,x^i).
\eea
And then
\begin{eqnarray}
  c_{\Delta,J}(R_A)
   &=& \lim_{x^i\to \infty}\frac{\langle{O}_{00\dots0}(0,x^i)\rangle_{\mathcal{C}^n_A} |x^i-x^i_A|^{2\Delta}}{N_{\Delta,J}a^{\nu_1\nu_2\dots\nu_J}Q_{00\dots0,\nu_1\nu_2\dots\nu_J}(n^\mu_s)}\nn\\
      &=&\frac{\langle\widetilde{{O}}_{00\dots0}(0)\rangle_{\widetilde{\mathcal{C}}^n_A} }{N_{\Delta,J}a^{\nu_1\nu_2\dots\nu_J}Q_{00\dots0,\nu_1\nu_2\dots\nu_J}(n^\mu_s)}.
\end{eqnarray}

Before proceeding, let us understand $\langle\widetilde{{O}}_{\Delta,J}(0)\rangle_{\widetilde{\mathcal{C}}^n_A}$ better. The distance between the point $(0,0,0,\ldots)$ and the conical singularity at $x^1=1/(2R_A)$ in $\widetilde{\mathcal{C}}^n_A$ is $1/(2R_A)$. In the following we make a global translation to shift the conical singularity to $(t_E=0,x^1=0)$ and the operator to $x^1=-1/(2R_A)$. Then we get
\be
 \langle\widetilde{{O}}_{\Delta,J}(0)\rangle_{\widetilde{\mathcal{C}}^n_A}=\langle\widetilde{{O}}_{\Delta,J}(0,-1/(2R_A),0,0,\ldots)\rangle_n
 \ee
where $n$ denotes the planar conical geometry with the singularity being located at $(t_E=0,x^1=0)$. Using the relation
\begin{equation}
\langle\widetilde{{O}}_{\Delta,J}(0,-1/(2R_A),0,0,\ldots)\rangle_n=(2R_A)^\Delta\langle\widetilde{{O}}_{\Delta,J}(0,-1,0,0,\ldots)\rangle_n,
\end{equation}
we have
\bea
c_{\Delta,J}(R_A)=\frac{(2R_A)^\D\langle\widetilde{{O}}_{00\dots0}(0,-1,0,\dots,0)\rangle_{n} }{N_{\Delta,J}a^{\nu_1\nu_2\dots\nu_J}Q_{00\dots0,\nu_1\nu_2\dots\nu_J}(n^\mu_s)}.
\eea
Plugging this into the expression of $R^0_{\Delta,J}$, we can get
\begin{eqnarray}
   R^0_{\Delta,J}&=&\frac{(2R_A)^\D\langle\widetilde{{O}}_{00\dots0}(0,-1,0,\dots,0)\rangle_{n} }{N_{\Delta,J}a^{\mu_1\mu_2\dots\mu_J}Q_{00\dots0,\mu_1\mu_2\dots\mu_J}(n^\mu_s)}
   \frac{(2R_B)^\D\langle\widetilde{{O}}_{00\dots0}(0,-1,0,\dots,0)\rangle_{n} }{N_{\Delta,J}a^{\nu_1\nu_2\dots\nu_J}Q_{00\dots0,\nu_1\nu_2\dots\nu_J}(n^\mu_s)}\nn\\
   &\times&a^{\mu_1\mu_2\dots\mu_J}a^{\nu_1\nu_2\dots\nu_J}
N_{\Delta,J}\frac{Q_{\mu_1\mu_2\dots\mu_J,\nu_1\nu_2\dots\nu_J}(n^\mu_s)}{|x_A-x_B|^{2\Delta}}.
\end{eqnarray}
As the one-point function in the conical geometry  is of the form
\bea
\langle\widetilde{{O}}_{\mu_1\mu_2\dots\mu_J}(x)\rangle_n=a_{\D,J}\frac{B_{\mu_1\mu_2\dots\mu_J}(n^\mu_x)}{r^\D},
\eea
where $r$ is the distance between $x$ and the singularity,  we have
\bea
\langle\widetilde{{O}}_{00\dots0}(0,-1,0,\dots,0)\rangle_n=a_{\D,J}B_{00\dots0}(n^\mu_1),
\eea
where $n^\mu_1$ is determined by $(0,-1,0,\dots,0)$. Then we find that
\bea
R^0_{\Delta,J}
&=&\frac{B^2_{00\dots0}(n^\mu_1)a^{\mu_1\mu_2\dots\mu_J}a^{\nu_1\nu_2\dots\nu_J}Q_{\mu_1\mu_2\dots\mu_J,\nu_1\nu_2\dots\nu_J}(n^\mu_s)}
{(a^{\mu_1\mu_2\dots\mu_J}Q_{00\dots0,\mu_1\mu_2\dots\mu_J}(n^\mu_s))^2}\frac{a^2_{\D,J}}{N_{\D,J}}\Big(\frac{4R_AR_B}{(x_A-x_B)^{2}}\Big)^\D\nn\\
&=&\frac{B^2_{00\dots0}(n^\mu_1)a^{\mu_1\mu_2\dots\mu_J}a^{\nu_1\nu_2\dots\nu_J}Q_{\mu_1\mu_2\dots\mu_J,\nu_1\nu_2\dots\nu_J}(n^\mu_s)}
{(a^{\mu_1\mu_2\dots\mu_J}Q_{00\dots0,\mu_1\mu_2\dots\mu_J}(n^\mu_s))^2}\frac{a^2_{\D,J}}{N_{\D,J}}z^\D
\eea

On the other hand, we have
\begin{equation}
  R_{\Delta,J}=s_{\Delta,J}G_{\Delta,J}(u,v).
\end{equation}
As the conformal block $G_{\Delta,J}(u,v)$  behaves as
\begin{equation}
  G_{\Delta,J}(u,v)=z^\Delta+\ldots\;\;\;,
\end{equation}
 we read
\begin{equation}
  s_{\Delta,J}=\frac{B^2_{00\dots0}(n^\mu_1)a^{\mu_1\mu_2\dots\mu_J}a^{\nu_1\nu_2\dots\nu_J}Q_{\mu_1\mu_2\dots\mu_J,\nu_1\nu_2\dots\nu_J}(n^\mu_s)}
{(a^{\mu_1\mu_2\dots\mu_J}Q_{00\dots0,\mu_1\mu_2\dots\mu_J}(n^\mu_s))^2}\frac{a^2_{\D,J}}{N_{\D,J}}.
\end{equation}
We can read out that
\begin{equation}
  f_{\Delta,J}=\frac{B^2_{00\dots0}(n^\mu_1)a^{\mu_1\mu_2\dots\mu_J}a^{\nu_1\nu_2\dots\nu_J}Q_{\mu_1\mu_2\dots\mu_J,\nu_1\nu_2\dots\nu_J}(n^\mu_s)}
{(a^{\mu_1\mu_2\dots\mu_J}Q_{00\dots0,\mu_1\mu_2\dots\mu_J}(n^\mu_s))^2}.
\end{equation}
Recalling that $n^\mu_s=(0,1,0,\dots,0)$ and $n^\mu_1$ is determined by $(0,-1,0,\dots,0)$,  we notice that $f_{\Delta,J}$ is independent of $x_A$ and $x_B$, and is just a constant.

In practice, we should first determine the form of $a^{\mu_1\mu_2\dots\mu_J}$. Then the computation is straight forward.
For the symmetric and traceless tensor, we find that
\bea
\frac{a^{\mu_1\mu_2\dots\mu_J}a^{\nu_1\nu_2\dots\nu_J}Q_{\mu_1\mu_2\dots\mu_J,\nu_1\nu_2\dots\nu_J}(n^\mu_s)}
{(a^{\mu_1\mu_2\dots\mu_J}Q_{00\dots0,\mu_1\mu_2\dots\mu_J}(n^\mu_s))^2}
&=&Q^{-1}_{00\dots0,00\dots0}(n^\mu_s),
\eea
and then
\bea
f_{\Delta,J}=\frac{B^2_{00\dots0}(n^\mu_1)}
{Q_{00\dots0,00\dots0}(n^\mu_s)}.
\eea
For the spin from $0$ to $4$, we find
\bea
B=1,& Q=1, &f_{\D,0}=1, \\
B_0=1, &Q_{0,0}=1,& f_{\D,1}=1, \\
B_{00}=1-d,&Q_{00,00}=1-\frac{1}{d},&f_{\D,2}=d(d-1), \\
B_{000}=1-d,&Q_{000,000}=1-\frac{3}{d+2},&f_{\D,3}=(d-1)(d+2),\\
B_{0000}=d^2-1,&Q_{0000,0000}=1-\frac{6}{d+4}+\frac{3}{(d+2)(d+4)},&
f_{\D,4}=(d-1)(d+1)(d+2)(d+4).\nn
\eea

\section{Numerical analysis}\label{numerical}

For free theories, it is more convenient to use real time approach to do numerical analysis. The ground state wave function can be obtained explicitly as the path integral is essentially Gaussian. The reduced density matrix of a region $A$ can be found by tracing out degrees of freedom outside region $A$. 
Then the entanglement entropy  of the region $A$ are \cite{Casini:2009sr}
\be
S_{A}=\tr((C_B+\frac{1}{2})\log(C_B+\frac{1}{2})-(C_B-\frac{1}{2})\log(C_B-\frac{1}{2}))
\ee
for free boson and
\be
S_A=-\tr((1-C_F)\log(1-C_F)+C_F\log C_F)
\ee
for free fermion.
To apply it to numerical analysis, we should use a discrete version of the $C_B$ and $C_F$ and find their eigenvalues.  It turns out that the eigenvalues of $C_B$ and $C_F$ can be fixed by the correlation functions inside region $A$. The matrix $C_B$ is
\bea
C_B&=&\frac{1}{2}\sqrt{1+\Lambda},\nn\\
\Lambda_{\vec{m}\vec{n}}&=&M_{\vec{m}\vec{k}}W_{\vec{k}\vec{n}}-\delta_{\vec{m}\vec{n}},\nn\eea
where the matrices $M$ and $W$ are respectively
\bea
W_{\vec{m}\vec{n}}&=&\int_{-\pi}^{\pi}\frac{d^{d-1}\vec{q}}{(2\pi)^{d-1}}e^{i\vec{q}\cdot(\vec{n}-\vec{m})}(2(\sum_{k=1}^{d-1}(1-\cos q_k)))^{\frac{1}{2}},\nn\\
M_{\vec{m}\vec{n}}&=&\int_{-\pi}^{\pi}\frac{d^{d-1}\vec{q}}{(2\pi)^{d-1}}e^{i\vec{q}\cdot(\vec{n}-\vec{m})}(2(\sum_{k=1}^{d-1}(1-\cos q_k)))^{-\frac{1}{2}}.\nn\eea
The matrix $C_F$ is of the form
\be
(C_F)_{\vec{m}\vec{n}}=<\psi^{\dag}_{\vec{m}}\psi_{\vec{n}}>=\frac{1}{2}\delta_{\vec{m}\vec{n}}+\frac{1}{2}\int_{-\pi}^{\pi}\frac{d^{d-1}\vec{q}}{(2\pi)^{d-1}}e^{i\vec{q}\cdot(\vec{m}-\vec{n})}\frac{\sum_{k=1}^{d-1}\sin q_k \gamma^0\gamma^k}{\sqrt{\sum_{k=1}^{d-1}\sin^2q_k}}. \nn
\ee
Here the matrix indices $\vec{m},\vec{n}$ are $d-1$ dimensional vectors, they label the spatial position of the lattice points.

To treat the matrices $W$ and $M$ more efficiently, we define a general integral
\be
W_{\vec{a}}^{\alpha}=\int_{-\pi}^{\pi}\frac{d^dq}{(2\pi)^d}e^{i\vec{q}\cdot \vec{a}}(2(d-\sum_{i=1}^d\cos{q_i}))^{\alpha},
\ee
in terms of which
\bea
W_{\vec{m}\vec{n}}&=&W^{\alpha}_{\vec{a}}(\alpha\to\frac{1}{2},\vec{a}\to \vec{n}-\vec{m},d\to d-1),\nn\\
M_{\vec{m}\vec{n}}&=&W^{\alpha}_{\vec{a}}(\alpha\to-\frac{1}{2},\vec{a}\to \vec{n}-\vec{m},d\to d-1).
\eea
Next we develop a method to compute $W_{\vec{a}}^{\alpha}$ for general $a_i\in \mathcal{Z}$ and $-1<\alpha<1$. Without losing generality, we can also assume $\alpha\not=0$.
\bea
W_{\vec{a}}^{\alpha}&=&(2d)^{\alpha}\int_{-\pi}^{\pi}\frac{d^dq}{(2\pi)^d}\prod_{i=1}^d\cos{a_i q_i}(1-\frac{\sum_{i=1}^d\cos{q_i}}{d})^{\alpha}\nn\\
&=&(2d)^{\alpha}\int_{-\pi}^{\pi}\frac{d^dq}{(2\pi)^d}\prod_{i=1}^d\cos{a_i q_i}\sum_{k=0}^{\infty}\frac{\Gamma(k-\alpha)}{\Gamma(-\alpha)k!d^k}(\sum_{i=1}^d\cos{q_i})^k\nn\\
&=&(2d)^{\alpha}\sum_{k=0}^{\infty}\frac{\Gamma(k-\alpha)}{\Gamma(-\alpha)k!d^k}\int_{-\pi}^{\pi}\frac{d^dq}{(2\pi)^d}\prod_{i=1}^d\cos{a_i q_i}\sum_{i=1}^d\sum_{k_i=0}^{\infty}\frac{k!}{k_1!k_2!\cdots k_d!}\prod_{i=1}^d(\cos{q_i})^{k_1}\delta_{k_1+\cdots+k_d,k}\nn\\&=&(2d)^{\alpha}\sum_{i=1}^d\sum_{k_i=0}^{\infty}\frac{1}{k_1!\cdots k_d!}\frac{\Gamma(k_1+\cdots +k_d-\alpha)}{\Gamma(-\alpha)d^{k_1+\cdots+k_d}}\prod_{i=1}^df(k_i,a_i)\nn
\eea
where we have defined
\be
f(k,a)\equiv\int_{-\pi}^{\pi}\frac{dq}{2\pi}\cos(aq)\cos^kq,\ k,a\in\{0\}\cup\{\mathcal{Z}^+\}.
\ee
The function $f(k,a)$ can be evaluated to be
\be
f(k,a)=\left\{\begin{array}{ll}
0,&\hs{3ex}k< a\\
\frac{(1+(-1)^{k+a})2^{-1-k}}{(1+k)B(\frac{2+k+a}{2},\frac{2+k-a}{2})},&\hs{3ex}k\geq a
\end{array}\right.
\ee
where $B(a,b)$ is the Euler Beta function.  Furthermore, we may use the integral representation of the Gamma function
\be
\Gamma(x)=\int_0^{\infty}ds s^{x-1}e^{-s},
\ee
and the following identity
\be
I_{a}(x)=\sum_{k=0}^{\infty}\frac{1}{(k+a)!}{x^{k+a}}{f(k+a,a)}
\ee
where $I_n(x)$ is the modified Bessel functions of the first kind, then we find that
 \be
 W_{\vec{a}}^{\alpha}=\frac{(2d)^{\alpha}}{\Gamma(-\alpha)}\int_0^{\infty}ds s^{-\alpha-1}e^{-s}\prod_{i=1}^d I_{a_i}(\frac{s}{d}).\label{w}
 \ee
Using the general expression (\ref{w}), we get
\bea
W_{\vec{m}\vec{n}}&=&\frac{\sqrt{2(d-1)}}{\Gamma(-\frac{1}{2})}\int_0^{\infty}ds s^{-\frac{3}{2}}e^{-s}\prod_{i=1}^{d-1} I_{|(\vec{n}-\vec{m})_i|}(\frac{s}{d-1})\label{wmn}\\
M_{\vec{m}\vec{n}}&=&\frac{1}{\sqrt{2(d-1)}\Gamma(\frac{1}{2})}\int_0^{\infty}ds s^{-\frac{1}{2}}e^{-s}\prod_{i=1}^{d-1} I_{|(\vec{n}-\vec{m})_i|}(\frac{s}{d-1})\label{invwmn}
\eea
where $|\vec{n}-\vec{m}|_i$ is the absolute value of $i$-th component of the vector $(\vec{n}-\vec{m})$.
In numerical computation, (\ref{wmn},\ref{invwmn}) are much easier to compute. There is one point which should be careful. As the integral of (\ref{wmn}) is not convergent 
  for $\vec{n}-\vec{m}=0$, we should use original integral to find the numerical value for the diagonal component of $W_{\vec{m}\vec{n}}$. 

For the free fermion, in deriving the matrix $C_F$, we may use the following identity to simplify the numerical computation:
\bea
J_{\vec{a}}^{\alpha}&=&\frac{1}{2}\int \frac{d^dq}{(2\pi)^d}\prod_{i=1}^d\cos a_iq_i(\sum_{i=1}^d\sin q_i^2)^{\alpha}\nn\\
&=&\left\{\begin{array}{ll}\frac{d^{\alpha}}{2^{\alpha+1}\Gamma(-\alpha)}\int_0^{\infty}ds e^{-s}s^{-\alpha-1}\prod_{i=1}^d I_{\frac{|a_i|}{2}}(\frac{s}{d}),& \forall a_i=0\ mod\ 2,\\
0,& \exists a_i=1\ mod\ 2.\end{array}\right.
\eea
There is one subtle point for the fermionic theory.  Once we use the lattice method to compute the entanglement entropy, we have doubled the degrees of freedom in each spatial direction. So the actual entanglement entropy is the numerical result divided by a factor $2^{d-1}$ in $d$ dimensions.

\end{document}